%% file: main.tex
\documentclass[11pt]{article}
\usepackage[paper=a4paper,margin=1.2in]{geometry} 

\usepackage{acro}
\usepackage[export]{adjustbox}
\usepackage{afterpage}
\usepackage{amssymb}
\usepackage{amsthm}
\usepackage{amsmath}
\usepackage{booktabs}              
\usepackage{capt-of}               
\usepackage{chemformula}
\usepackage{comment}               
\usepackage{crimson}
\usepackage{draftfigure}
\usepackage[official]{eurosym}
\usepackage{float}
\usepackage[hidelinks]{hyperref}
\usepackage[utf8]{inputenc}
\usepackage{multirow}
\usepackage{setspace}
\usepackage{soul}                  
\usepackage{subcaption}
\usepackage{tabularx}
\usepackage{tabulary}
\usepackage{varwidth}
\usepackage{xcolor}

\newsavebox\tmpbox 

\usepackage[
backend      = biber,
natbib       = true,
citestyle    = authoryear,
bibstyle     = authoryear,
date         = year,
url          = true,
urldate      = long,
isbn         = false,
doi          = true,
maxcitenames = 1,
maxbibnames  = 9,
uniquelist   = true
]{biblatex}

\DeclareFieldFormat*{url}{}
\DeclareFieldFormat[online]{url}{\mkbibacro{URL}\addcolon\space\url{#1}}
\DeclareFieldFormat[report]{url}{\mkbibacro{URL}\addcolon\space\url{#1}}
\DeclareFieldFormat*{urldate}{}
\DeclareFieldFormat[online]{urldate}{\mkbibparens{\bibstring{urlseen}\space#1}}

\DeclareFieldFormat[article,misc,inbook,book,report,online]{title}{\mkbibquote{#1\isdot}} 

\renewbibmacro{in:}{}

\DeclareAcronym{ashp}{
  short = ASHP,
  long = air-sourced heat pump,
  long-plural = s,
}

\DeclareAcronym{cop}{
  short=COP,
  long=coefficient of performance
}

\DeclareAcronym{eraa}{
  short=ERAA,
  long=European Resource Adequacy Assessment
}

\DeclareAcronym{gw}{
  short=GW,
  long=Gigawatt,
}

\DeclareAcronym{ghg}{
  short=GHG,
  long=greenhouse gas,
}

\DeclareAcronym{ntc}{
  short=NTC,
  long=net-transfer capacity,
}

\DeclareAcronym{p2g2p}{
  short=p2g2p,
  long=power-to-gas-to-power,
}

\DeclareAcronym{pv}{
  short=PV,
  long=solar photovoltaic,
}

\DeclareAcronym{twh}{
  short=TWh,
  long=Terawatt hour,
  short-plural = ,
  long-plural = s,
}

\DeclareAcronym{us}{
  short=US,
  long=United States
}

\title{Power sector impacts of a simultaneous European heat pump rollout}

\author{Alexander Roth\thanks{\href{mailto:aroth@diw.de}{aroth@diw.de}}\\ \small{German Institute of Economic Research (DIW Berlin) \& TU Berlin}}

\date{December 11, 2023}

\begin{document}


\maketitle
\thispagestyle{empty}


\begin{abstract}

\noindent The decarbonization of buildings requires the phase-out of fossil fuel heating systems. Heat pumps are considered a crucial technology to supply a substantial part of heating energy for buildings. Yet, their introduction is not without challenges, as heat pumps generate additional electricity demand as well as peak loads. To better understand these challenges, an ambitious simultaneous heat pump rollout in several central European countries with an hourly-resolved capacity expansion model of the power sector is studied. I assess the structure of hours and periods of peak heat demands and their concurrence with hours and periods of peak residual load. In a 2030 scenario, I find that meeting 25\% of total heat demand in buildings with heat pumps would be covered best with additional wind power generation capacities. I also identify the important role of small thermal energy storage that could reduce the need for additional firm generation capacity. Due to the co-occurrence of heat demand, interconnection between countries does not substantially reduce the additional generation capacities needed for heat pump deployment. Based on six different weather years, my analysis cautions against relying on results based on a single weather year.

\end{abstract}

\bigskip
\bigskip


\newpage


\onehalfspacing

\section{Introduction}

\begin{figure}[!htb]
    \centering
        \begin{subfigure}{.66\textwidth}\centering\includegraphics[width=\textwidth]{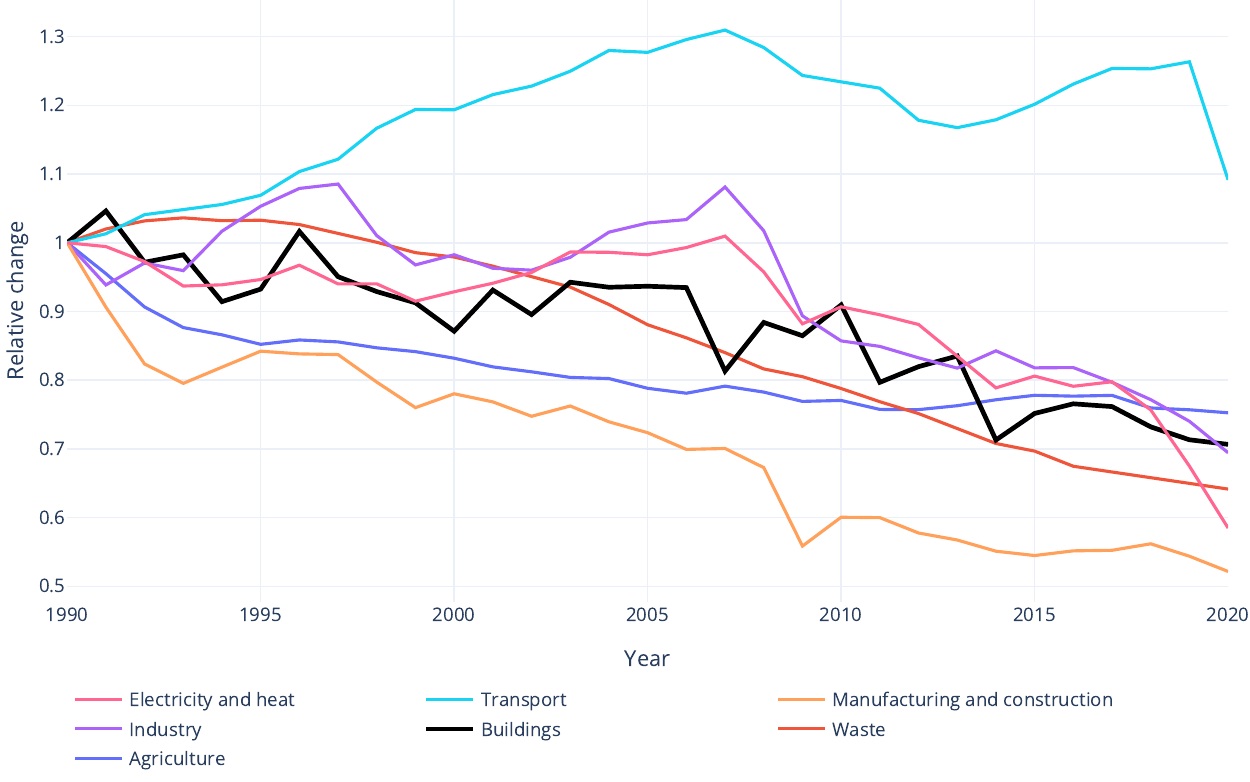}\subcaption[]{Relative development (1990 = 1)}\label{fig:emissions_relative}\end{subfigure}%
        \begin{subfigure}{.33\textwidth}\centering\includegraphics[width=\textwidth]{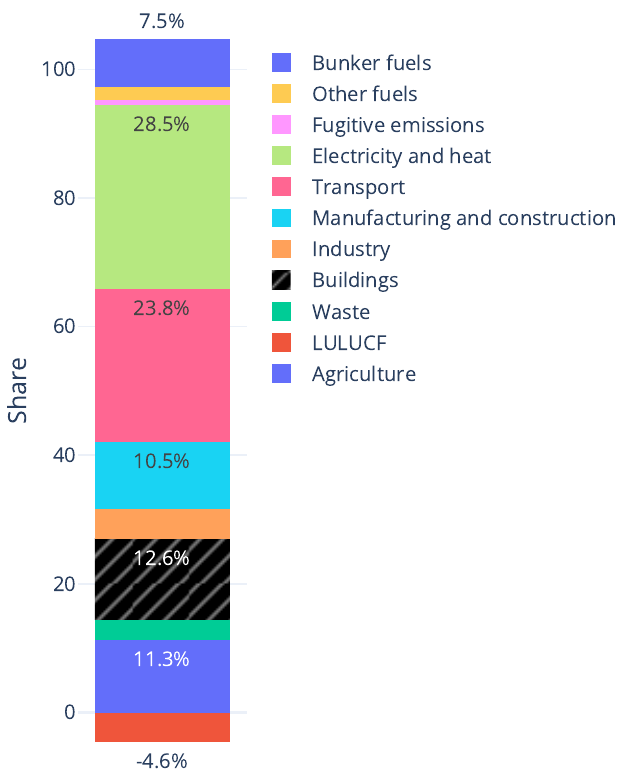}\subcaption[]{Shares in 2019}\label{fig:emissions_bar}\end{subfigure}%
    \vspace{1mm}
    \begin{minipage}[c]{0.99\textwidth}
    \medskip\footnotesize
        \emph{Source:} Own illustration based on \citet{ritchie_co2_2023}.
    \end{minipage}
    \caption{Sectoral \ac{ghg} emissions in the EU 27}
    \label{fig:emissions}
\end{figure}

To limit the increase of global mean temperature and mitigate its consequences, European countries have decided to decrease their \ac{ghg} emissions in the coming decades, achieving a net-zero economy in 2050 \citep{europeanclimatelaw_regulation_2021}. Reaching this goal requires decreasing \ac{ghg} emissions in all sectors of the economy. The building sector, mainly heating and cooling, contributes significantly to carbon emissions (Figure \ref{fig:emissions_bar}). Despite some progress in recent years (Figure \ref{fig:emissions_relative}), further reductions are needed, especially in large countries such as Germany, which still rely heavily on fossil fuel heating systems. 
One solution to decrease \ac{ghg} emissions in the building sector is to deploy non-fossil heating technologies, such as heat pumps. This study focuses on a simultaneous and substantial rollout of heat pumps in several central European countries and assesses the challenges for the power sector.

As the building sector is responsible for around 13\% of total greenhouse gas emissions in the EU (Figure \ref{fig:emissions_bar}), the role of this sector in reducing its overall emissions is paramount.\footnote{Importantly, these reported emissions from the building sector only cover emissions directly emitted at the building, while emissions from the generation of electricity and heat in power plants are reported in \textit{Electricity and heat}. On top of that, emissions associated with the construction of buildings are also not contained. Hence, the overall share of building-related emissions is larger than shown in Figure \ref{fig:emissions_bar}.} Several technological options exist that are already mature and widely available to decarbonize the building sector \citep{climateactiontracker_decarbonising_2022}. On the demand side, insulation, for instance, can reduce the overall energy needs of buildings, while on the supply side, traditional fossil fuel heating solutions have to be replaced. Among others, heat pumps are considered a crucial technology to achieve decarbonization \citep{iea_future_2022}. A heat pump extracts heat from a source, such as the ambient air or the ground, and transports it to a destination where it is needed, such as a water-based heating system within a building. Most heat supplied by a heat pump is harvested from the environment, while the electricity is mainly used to transfer and lift the heat to a useful temperature level \citep{iea_future_2022}. Therefore, heat pumps possess two essential features: they are efficient and operate directly with electricity. Mainly due to the latter, their usage lends itself perfectly to a decarbonized energy system, relying almost exclusively on renewable electricity, in which heat pumps can be run directly without an intermediary energy carrier.

However, an ambitious deployment of heat pumps does not come without challenges. Apart from possible electricity transmission and distribution grid requirements, the direct use of electricity by heat pumps, mentioned above as an advantage, also poses challenges for the power sector. The rollout would not only lead to an increase in electricity demand - which would have to be covered by additional generation capacities - but could also lead to more elevated peak loads in the power sector, requiring additional flexibility. This flexibility could be attained, for instance, by firm generation capacities (fossil, renewable, or storage), interconnection between countries, or by thermal energy storage of heat pumps. As higher shares of renewable electricity already increase the need for flexibility options, heat pumps possibly add to this need.


Several studies assess the effects of heat pumps on the power sector. Early on, \citet{hedegaard_energy_2013} point out the benefits of flexible operation of heat pumps. In a recent working paper, \citet{roth_flexible_2023} analyze the additional generation capacities needed for different heat pump rollout speeds in Germany and the impact of thermal energy storage. However, only a deployment in Germany is considered, and capacity expansion in neighboring countries is not modeled. \citet{altermatt_replacing_2023} assess an even more ambitious heat pump rollout path for Germany, yet neither explicitly modeling the electricity sector nor accounting for other countries. Concerning the flexibility of heat pumps, \citet{kroger_electricity_2023}, using a detailed modeling approach for small- and large-scale heat pumps, quantify additional peak loads through heat pumps and shifting potentials through thermal energy storage. Yet, the heat pump expansion is limited to Germany, and no capacity expansion effects are estimated. In a case study of the British and Spanish market, \citet{lizana_national_2023} determine the optimal thermal energy storage size to shift peak power demand, yet neither explicitly modeling the electricity sector. With respect to additional peak loads generated by heat pumps, \citet{charitopoulos_impact_2023} claim that heat demand peaks are often considerably lower than values widely cited in the literature, that deep electrification of heating can be achieved with moderately higher electricity load peaks, and that thermal energy storage plays an important role in shifting loads. On the other hand, \citet{buonocore_inefficient_2022} conclude that a poorly executed electrification of heat in the \ac{us} would require a massive expansion of renewable energies. Using a regression-based approach in the UK, \citet{deakin_impacts_2021} estimate the additional peak demand of heat pumps, while \citet{chen_impact_2021} conclude that the electrification of heat is cost-effective compared to other solutions, but lead to considerable additional demand that has to be met - in their study - by wind energy installations. Finally, \citet{hilpert_effects_2020} highlight the importance of flexible heat pump operation in 100\% renewable energy systems, relating well to the findings of other studies.



















As highlighted above, most of the literature (1) assesses heat pump deployment only in single countries, (2) does not account for interconnection, or (3) considers only a single weather year. Hence, this study assesses a parallel and ambitious heat pump rollout in several countries and aims to identify its effects on the power sector in a midterm 2030 setting. While large-scale studies have simulated decarbonization pathways for Europe, including heat pumps, this study seeks to isolate the heat pump effect on the electricity system. I assess the required power plant additions, especially the firm capacity additions. I estimate the impact of an important source of flexibility: I study how a small thermal energy storage attached to the heat pumps influences the residual load and, therefore, generation capacity needs. I also assess how much flexibility can be provided by interconnection between countries. Crucially, I study how strongly cold spells are correlated in Europe and how they overlap with the residual load. I do not base my analysis on a single weather year but consider six different weather years. By relying on several years, I not only improve the stability of the results but also provide an intuition for the variability of heat demand and its impact in Europe. Finally, I run several robustness checks to assess the robustness of my results.

\section{Model and Data} \label{sec:data and modelling}

My analysis builds on several tools and data sources mentioned in this section. While the functioning of the overall power sector model is explained in Section \ref{sec:electricity sector}, the following subsection briefly explains the fundamentals of the heat module used.

\subsection{Model}

\subsubsection{Heating}

I use a straightforward approach to model the interaction between heat demand for space and water in buildings and the power sector. As explained in the subsequent subsection \ref{sec:data} on data, I use as a crucial input the heat demand for space and water for different house categories at every hour of the year and in every country. Exogenously, I assume a share $s_{bt,st,hpt}$ of which heat demand $hd_{bt,st,h}$ has to be covered by heat generated $HO_{bt,st,htp,h}$ by heat pumps in every hour $h$ of the year.\footnote{If not differently noted, all parameters, variables, and equations mentioned in this section apply equally to every country. For the reason of simplicity and readability, I omit a country-specific subscript.} As the heat demand is exogenously given, it is assumed to be totally inelastic. In the present model set-up, the model has to fulfill that heat demand and has no possibility of not serving it.

\begin{equation}
 HO_{hp,st,h} = s_{bt,st,hpt} \times hd_{bt,st,h},
\end{equation}

where $bt$ is the \textit{building type} (single-family, multifamily, or commercial), $st$ the type of heat sink (space or water), and $hpt$ the type of heat pump (air-sourced, ground-sourced, water-sourced).

As heat pumps can be equipped with thermal energy storage, the following equation governs the state of charge $HL$ of that storage: 

\begin{equation}
HL_{bt,st,hpt,h} = HL_{bt,st,hpt,h-1} + HI_{bt,st,hpt,h} - HO_{bt,st,hpt,h}.
\end{equation}

The state of charge $HL$ increases with heat supplied $HI$ and decreases with heat output $HO$. The required heat output $HO_{bt,st,hpt,h}$ of the heat pump can either be met using heat from thermal storage $HL$ or generating it $HI$. Obviously, the state of charge is always 0 ($HL = 0$) if heat pumps do not have thermal energy storage. In that case, heat output $HO$ and heat generated $HI$ are equal every hour. The size of the thermal storage, hence the maximum of $HL$, is determined by an exogenously set \textit{energy-to-power} ratio $ep_{bt,st,hpt}$ that relates the maximum heat output (the installed heat output capacity) to the size of the thermal energy storage.

To generate heat $HI_{bt,st,hpt,h}$, heat pumps use electricity $E_{bt,st,hpt,h}$. The sink-, heat pump-, and hour-specific \ac{cop} determines that process and, hence, the efficiency of the heat pump. The higher the \ac{cop}, the more heat is generated with the same electricity input:

\begin{equation}
HI_{bt,st,hpt,h} = cop_{st,hpt,h} \times E_{bt,st,hpt,h}.
\end{equation}

With the present model formulation, I treat space and water heating separately using different \ac{cop}s. The \ac{cop} for water is generally lower due to the higher temperature needed compared to space heating. While this formulation is probably slightly unrealistic for many houses with a single heat pump system to serve space and water heating, my specification determines the electricity needed more precisely. In any case, the difference to a model in which the entire heat of a house is served with the same \ac{cop} is not substantial, as overall heat demand for water is relatively small compared to space heat demand.

The installed capacity of the heat pumps, with respect to heat output, electricity input, and thermal energy storage, is not determined based on cost-optimally but is set to satisfy heat demand every hour. Therefore, its size is chosen to meet the peak heat demand, given the \ac{cop} of that hour, in the absence of thermal energy storage.

\subsubsection{Power sector} \label{sec:electricity sector}

To measure the impact of a heat pump rollout on the power sector, I use the electricity sector model DIETER \citep{zerrahn_longrun_2017,schill_longrun_2018,gaete-morales_dieterpy_2021}, which derives optimal dispatch and investment decisions. This model has been used in numerous peer-reviewed publications \citep[e.g.][]{gils_modelrelated_2022,roth_renewable_2023,kirchem_power_2023}. DIETER is a linear cost-minimization model that takes all 8760 consecutive hours of a year into account and optimizes investment and dispatch of the power sector. The model does not contain a detailed grid but assumes a ``copper plate'' within a country, while a \ac{ntc} model is used between countries. Important endogenous variables are the capacity installation of power plants and storage, the dispatch, as well as the power flow between countries. For a detailed model formulation, I refer to \citet{gaete-morales_dieterpy_2021}. For this analysis, the subsequent features and assumptions characterize the model. Unless differently stated, these hold for all model runs.

\paragraph{Generation} In terms of generation technologies, the following are present in this analysis: \textit{variable renewables}: \ac{pv}, onshore and offshore wind power, run-of-river hydropower; \textit{dispatchable renewables:} bioenergy, reservoir hydropower; \textit{non-renewables:} nuclear power, gas-fired power (closed-cycle turbine) (CCGT), lignite, hard coal, oil, other.

\paragraph{Storage} Lithium-ion batteries, \ac{p2g2p} storage, pumped-hydro storage: with inflow (open PHS) and without inflow (closed PHS).

\paragraph{Capacity bounds} In principle, capacity installations of the different generation and storage technologies are not restricted. However, to increase the realism of the scenarios, I restrict certain technologies with upper and lower bounds. Whenever I impose bounds, I use the values given by \citetitle{entso-e_eraa_2021}, using the year 2025 to take values as close as possible to current values (Table \ref{tab:capacity_bounds}). \ac{pv}, on- and offshore wind have no upper capacity bounds, but lower bounds are set according to \ac{eraa}. All hydro technologies (run-or-river, open and closed PHS, reservoir) are fixed. Gas power plants have lower bounds but are not restricted to their upper bounds. In this manner, I allow for the addition of capacity while avoiding (unlikely) decommissioning until 2030. Similarly, I fix the values of hard coal and lignite power plants to account for the existing fleet that will not be further expanded but is likely to stay in operation as a backup. The capacities of nuclear power are also fixed, with the idea in mind that changes (additions or decommissioning) are unlikely until 2030. All remaining technologies (oil, other) have upper bounds, yet no lower bounds. The power and energy capacities of battery and hydrogen storage are unconstrained. Please note that some of these assumption are altered in the robustness checks. All capacity bounds are shown in Table \ref{tab:capacity_bounds}.

\paragraph{Generation bounds} No bounds for yearly total generation are set for any technology except for bioenergy, for which I constrain generation to 2022 values \citep{ember_yearly_2023}.

\paragraph{Net-transfer capacities} The net-transfer capacities between countries are fixed exogenously and are based on \citetitle{entso-e_eraa_2021}, using the year 2025.

\paragraph{Renewable electricity share} No minimum share of renewable electricity production on total electricity production or consumption is assumed.

\paragraph{\ch{CO2} price} A price of 150 \EUR{} per ton of \ch{CO2} emitted is assumed.

\paragraph{Countries} The scenarios entail the following nine countries: Austria, Belgium, Denmark, France, Germany, Italy, Luxembourg, Netherlands, and Switzerland\footnote{\label{note1}As heat demand data for Switzerland is missing, Switzerland is part of the analysis and optimization, yet not heat pump rollout is simulated there.}.

\subsection{Data} \label{sec:data}

\paragraph{Technology and costs} For the technology and cost data, I rely primarily on \citet{gaete-morales_dieterpy_2021}. Table \ref{tab: costs} shows the values.

\paragraph{Heating demand time series} Relying on the \textit{When2Heat} dataset \citep{ruhnau_time_2019} and its latest update and extension \citep{ruhnau_update_2022}, I use the total national space and water gas heat demand. The database provides hourly profiles of heat demand differentiated between different building types (single family, multifamily, commercial) and sink (space and water), and hourly \ac{cop}s for different types of heat pumps (air-sourced, ground-sourced, ground-water-sourced), separated for sinks. The heat demand data are available for the years 2009–2015.\footref{note1}

\paragraph{Renewable availability and electricity demand time series} The data provided by \citeauthor{entso-e_eraa_2021}, used in \citetitle{entso-e_eraa_2021}, has time series of renewable availability, hydro inflows \textit{(Pan-European Climate Database)}, and electricity demand for different weather years of all European countries. Specifically, I rely on the machine-friendly version of \citet{defelice_entsoe_2022}. 

\section{Scenarios} \label{sec:scenarios}

Given the features and assumptions in the previous section, I model the heat pump deployment by requiring a certain share, 25\%, of total heat demand to be covered by heat pumps. Please note that for simplicity, I assume the same share for single-family houses, multiple-family houses, and commercial buildings, as well as space and water. I also consider only one type of heat pump, air-sourced, to cover the heat, with the idea that \acp{ashp} currently dominate the market. It is important to mention that I do not model heat demand in a strict bottom-up way. Therefore, I do not assume anything about the part of existing electricity demand used to generate heat or the remaining heat demand not covered by the model. Therefore, the model and scenario definitions implicitly assume that I only assess the effect of \textit{additional} heat pumps that cover 25\% of total building heat demand \textit{additionally}. 

Table \ref{tab:scenarios} provides an overview of the definition of my \textit{base} scenarios. As a reference, I conduct a scenario run in which no heat has to be covered by heat pumps. In the two other scenarios, 25\% of the heat has to be covered by heat pumps with varying sizes of thermal energy storage. In one scenario, all heat pumps are equipped with thermal energy storage sized at two hours of the maximum heat output. In the other scenario, this size is zero; hence, no thermal energy storage is available. I assume that heat pump owners are faced with wholesale electricity prices and operate their heat pumps in a system-friendly way. The advantage of thermal energy storage is the possibility of moving electric demand induced by heat demand to hours, in which the residual load (total load minus the generation of variable renewable electricity) is lower; hence, prices are lower. Another advantage of moving heat pump load away from hours of heat demand (and likely lower temperatures) is that heat pumps can possibly generate heat in hours of higher temperatures, therefore higher \ac{cop}s, lowering the overall electricity demand of heat pumps.

Every scenario run is conducted for six weather years (2009-2014). To adequately capture the heating period in each year, I do not run the model from January to December, as it is commonly done in energy system modeling, but from July to June. If a specific weather year is mentioned in the following, I refer to the period starting in July and ending in June of the following year. For instance, the year 2009 would refer to the period July 2009 to June 2010.

\begin{table}[!htb]
    \centering
     \caption{Definition of \textit{base} scenarios}
     \label{tab:scenarios}
    \begin{tabular}{c|c}
    \toprule
         Heat share &  E/P ratio of thermal storage \\ \midrule
         0\%& -\\
        25\%& 0\\
        25\%& 2\\
 \bottomrule
 \end{tabular}
\end{table}

To check the robustness of my results, additional scenario runs are employed, in which specific assumptions are varied (Table \ref{tab:robustness checks}). For all robustness checks, a scenario with no heat pumps and a scenario with 25\% heat covered by heat pumps is conducted. In all runs, heat pumps are equipped with a two-hour thermal heat storage. All scenarios are run for six weather years. 

The scenario \textit{gas\_free} removes the lower bounds of gas-fired power plants, checking whether the model would prefer to install less CCGT capacity. \textit{no\_nuc} assumes a nuclear power plant fleet that is 50\% lower than current values, assessing the impact of a partial nuclear phase-out. \textit{no\_coal} assumes a total decommissioning of lignite and hard coal-fired power plants. \textit{no\_ntc} is a counterfactual scenario in which no power flows between countries are possible, estimating the importance of cross-border electricity trade. As wind power might face expansion restrictions, \textit{wind\_cap} is a scenario in which wind on- and offshore capacities can only be expanded by 50\% beyond ERAA 2021 values.

\begin{table}[!htb]
    \centering
    \caption{Robustness checks}
    \label{tab:robustness checks}
    \settowidth\tymin{\textbf{Scenario}}

        \begin{tabulary}{\textwidth}{L|L}
        \toprule
            Scenario &Description\\ \midrule
            \textit{gas\_free} & No capacity (lower or upper) for gas-fired power plants.  \\
            \textit{half\_nuc} & Nuclear power plant capacities fixed at 50\% lower value compared to \textit{base}. \\
            \textit{no\_coal}  & No hard coal or lignite power plants.                     \\
            \textit{no\_ntc}   & No electricity flow between countries.             \\
            \textit{wind\_cap} & Upper bounds for on- and offshore wind power capacity at 50\% above ERAA 2021 values. \\ \bottomrule
        \end{tabulary}
\end{table}

\section{Results \& Discussion}

Before showing and discussing the principal model outcomes of the scenarios, a few fundamental facts about heat pumps and heat demand regarding the scenarios are presented; then, hours and periods of peak heat demand are analyzed; and, finally, the main outcomes, primarily generation capacities, are shown.

\begin{figure}[!htb]
\centering
\begin{minipage}{0.495\textwidth}
\begin{table}[H]
    \centering
        \resizebox{1.1\linewidth}{!}{\input{tables/table_cap}}
    \caption{Heat pump capacities}
    \label{tab:heat capacities}
\end{table}
\end{minipage}
\hfill
\begin{minipage}{0.495\textwidth}
\begin{figure}[H]
    \centering
    \includegraphics[scale=0.33]{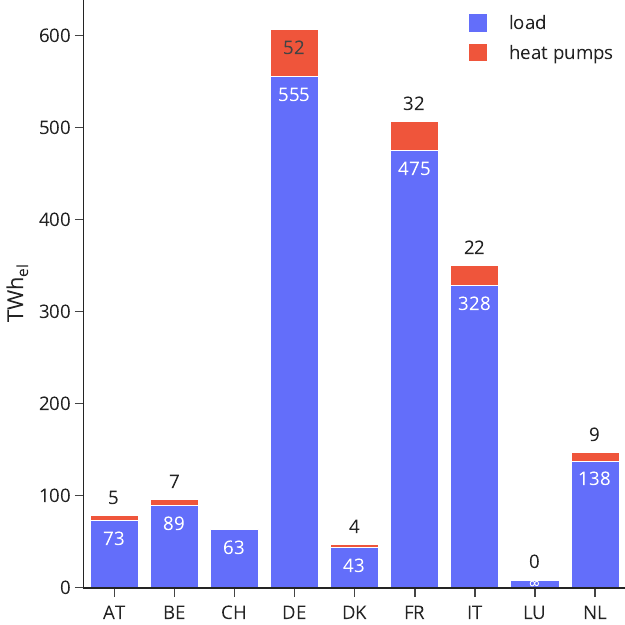}
    \includegraphics[scale=0.33]{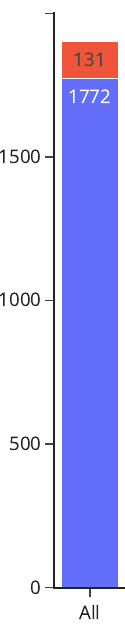}
    \caption{Yearly electricity demand} 
    \label{fig:electricity_demand}
\end{figure}
\end{minipage}\par%
    \vspace*{0.3cm}
    \begin{minipage}{\textwidth}
        \footnotesize \textit{Note:} The values depicted result from the \textit{base} scenario with thermal energy storage of two hours and the weather year 2009.
    \end{minipage}
    \vspace{0.3cm}
\end{figure}

In the \textit{base} scenario, 25\% of total space and water heating is supplied by \acp{ashp}, leading to a substantial electricity demand (Figure \ref{fig:electricity_demand}). For instance, Germany would need to cover around 52 \acp{twh} in addition to its already existing load of 555 \acp{twh}, roughly an increase of ten percent. As explained in section \ref{sec:data and modelling}, the size of the heat pumps in terms of electricity input, heat output, and thermal storage capacities are not determined endogenously but are set so that heat demand can be covered in every hour, even without thermal storage. With these assumptions in mind, Table \ref{tab:heat capacities} depicts the installed heat pump capacities that would follow the requirement to cover 25\% of total building heat demand with \acp{ashp}. Relevant for the electricity sector is the installed electricity input capacity of heat pumps, which would reach almost 40 \ac{gw} in Germany and over 20 \ac{gw} in France.

\subsection{Heat demand peaks and total heat demand do not align}

\begin{figure}[!htb]
    \centering
        \begin{subfigure}{.33\textwidth}\centering\includegraphics[width=\textwidth]{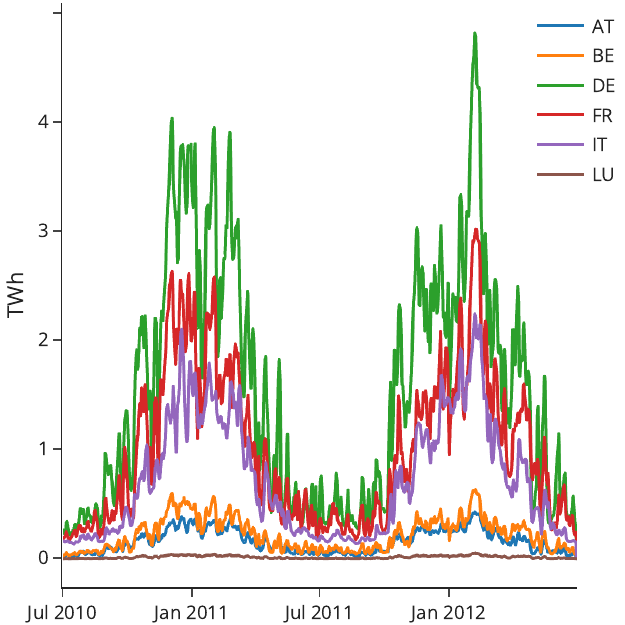}\subcaption[]{Daily}\label{fig:heat_demand_ts}\end{subfigure}%
        \begin{subfigure}{.33\textwidth}\centering\includegraphics[width=\textwidth]{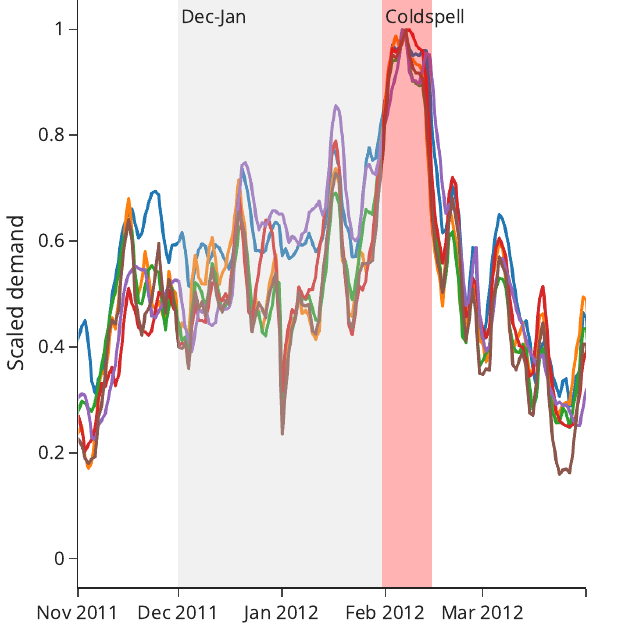}\subcaption[]{Daily scaled to max. per year}\label{fig:heat_demand_ts_scaled}\end{subfigure}%
        {\begin{subfigure}{.33\textwidth}\centering\includegraphics[width=\textwidth]{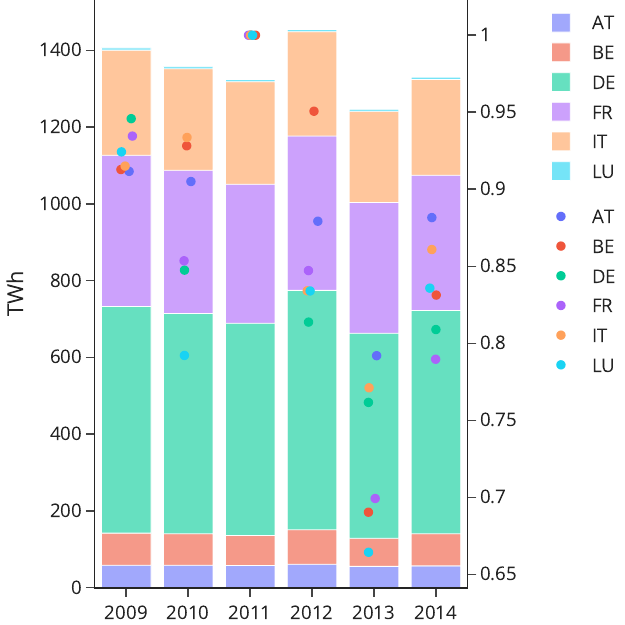}\subcaption[]{Yearly and max. hour}\label{fig:heat_demand_total}\end{subfigure}\par}
        \begin{minipage}{\textwidth}
            \medskip \footnotesize \textit{Note:} In panel (c), yearly heat demand is shown on the left axis. The maximum hourly heat demand, scaled to the overall maximum hourly heat demand of the entire period, is depicted on the right axis. 
        \end{minipage}
    \caption{Heat demand}
    \label{fig:heat_demand}
\end{figure}

As mentioned above, a simultaneous heat pump rollout could pose several challenges for the European electricity sector in terms of overall electricity needed and additional peak loads. Regarding these questions, a sensible approach is to first look at the overall structure and principal characteristics of heat demand for the countries included in my analysis (Figure \ref{fig:heat_demand}). Not surprisingly, heat demand increases in winter in all countries (Figure \ref{fig:heat_demand_ts}). The figure also reveals the dimensions of energy needed for residential and commercial heating of buildings: at the peak, almost five \acp{twh} of thermal energy were used per day for space and water heating in Germany in residential and commercial buildings in the year 2012. Despite the correlational appearance of heat demand, suggested by Figure \ref{fig:heat_demand_ts}, heat demand patterns are more nuanced. The winter of 2011-2012 serves as a good example, in which heat demand showed only a partial correlation in the months of December and January. Yet, a cold spell hit the continent in mid-February, and heating demand surged simultaneously. Scaled to its maximum yearly value, total daily heat demand peaked in all countries around the same time (Figure \ref{fig:heat_demand_ts_scaled}). Two insights for the energy system might arise: even a mild winter can be a strain if it contains a short cold spell, while moderately cold winters might be less of a challenge if they do not go beyond the expected. Related to this question, no clear relationship can be found between total heating energy needed and peak heating needed: Figure \ref{fig:heat_demand_total} shows the yearly (July-June) heating demand for each country in bars (left axis), while the dots depict the maximal hourly heat demand, scaled to the overall maximum heat demand of the period 2009-2014 (right axis). The hour of maximum heat demand in the period 2009-2014 occurred in all countries in 2011. In other years, the respective maximum heat hours show considerably lower values. Interestingly, 2011 was clearly not among the coldest years, as overall heat demand is lower than in years before and after. As peak heat demand was the highest, peak and total heat demand do not need to coincide.

\subsection{Heat pump load peaks do not necessarily align with residual load peaks} \label{sec:results heat}

\begin{figure}[H]
    \centering
        \begin{subfigure}{.33\textwidth}\centering\includegraphics[width=\textwidth]{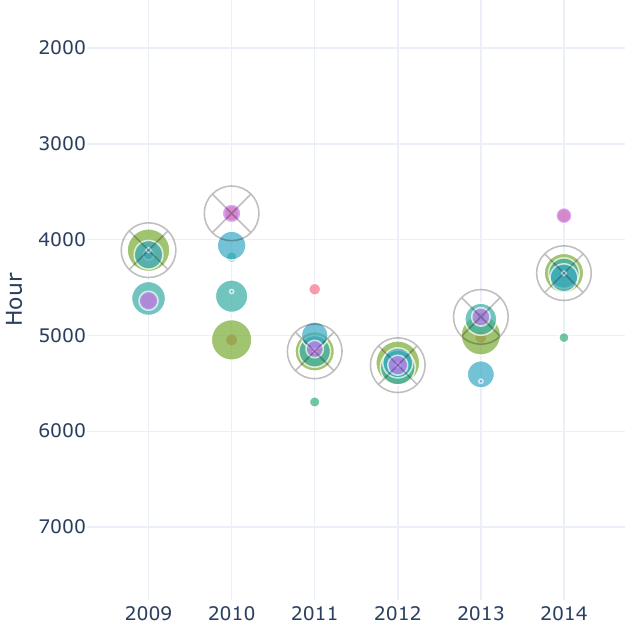}\subcaption[]{Maximum heat demand}\label{fig:max_heat_demand_hour}\end{subfigure}%
        \begin{subfigure}{.33\textwidth}\centering\includegraphics[width=\textwidth]{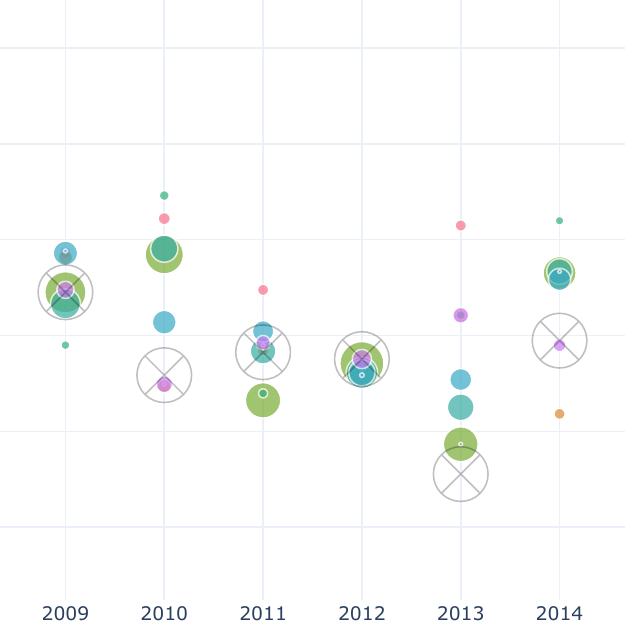}\subcaption[]{Maximum heat pump load}\label{fig:max_hp_load_hour}\end{subfigure}%
        \begin{subfigure}{.33\textwidth}\centering\includegraphics[width=\textwidth]{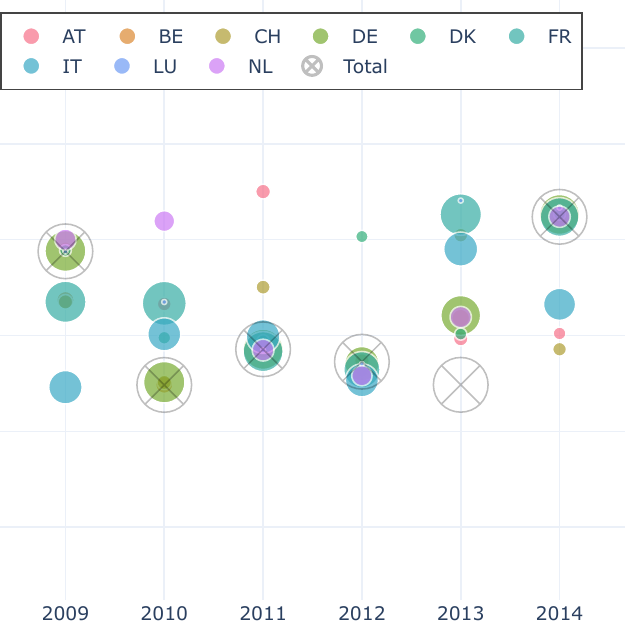}\subcaption[]{Maximum residual load}\label{fig:max_rl_hour}\end{subfigure}\par
        \begin{minipage}{\textwidth}
            \medskip \footnotesize \textit{Note:} The values depicted results from the \textit{base} scenario with thermal energy storage of two hours. \textit{Total} refers to the hour in which the sum of all countries is at its maximum. The size of that associated gray marker is not to scale. In panel (c), the residual load does not contain the electricity demand of heat pumps. \par
        \end{minipage}
    \caption{Maximum heat demand, heat pump load, and residual load: size and hour}
    \label{fig:max hours}
\end{figure}

However, it is important to acknowledge that the analysis of peak heat demand, even for several countries, does not fully capture the actual challenge for the electricity sector. While cold spells could lead to severe peak loads by heat pumps, consequences would be limited if these hours were accompanied by a high generation of renewable electricity. Therefore, it is helpful to analyze the relationship between the residual load, defined as electricity demand minus the generation of all variable renewable energy sources (photovoltaic, wind on- and offshore, and hydro run-of-river), and peak heat demand (Figure \ref{fig:max hours}): the occurrence of the hour of the maximum heat demand of every country in every year, as well as its relative size (indicated by the diameter of the circle) is depicted (\ref{fig:max_heat_demand_hour}), as well as the hour of maximum heat pump load (\ref{fig:max_hp_load_hour}), and the hour of the maximum residual load (\ref{fig:max_rl_hour}). The hour of maximum heat demand can be found in the winter months. Varying between the years, hours of maximum heat demand normally occur between the hours 4000 and 5500 (starting July 1st). In some years (like 2012), there is a coincidence of all maximum hours, while in other years (like 2010), no alignment can be seen. In many years, the hour in which the sum of heat demand of all countries is maximal also coincides with the maximum hour in the individual countries. As Figure \ref{fig:max hours} depicts values of a scenario in which heat pumps are equipped with thermal energy storage of two hours, the maximum heat pump load (\ref{fig:max_hp_load_hour}) does not necessarily coincide with the maximum heat demand (\ref{fig:max_heat_demand_hour}). In some countries and in some years, the maximum heat pump load and maximum heat demand align, such as in the year 2012. However, for many years, the maximum heat pump load is at a different hour than the maximum heat demand, suggesting that the model has used thermal energy storage to disentangle heat demand and heat pump lead. For the impact on the power sector, though, it is relevant to see whether the maximum heat pump load coincides with the maximum residual load. Only if they fall together, the power sector would be strained, and additional (firm) generation capacity would be needed to cover the load. Like heat demand, all maximum residual load events can be found in the winter period (\ref{fig:max_rl_hour}). The exact occurrences of the peak residual load hours vary quite strongly between years. Just like heat demand, they fall together in all countries in some years (2012), while they do not in others (2013). Most importantly, in many years and many countries, including the sum of all countries, maximum heat load and maximum residual do not align, suggesting a possible limited impact on the power sector. Yet, for the years 2011 and 2012, they align very well (roughly after the hour 5000), which should reflect a higher need for firm capacity in these years. The alignment could exist because wind speeds could be correlated with temperatures, and the existing electricity demand could already cover part of the heat demand, hence peaking in the same hour. It is also important to mention that the figure depicts only the hours of maximum heat demand, heat pump load, and residual load. However, the hours with the second, third, etc. highest values could show a different correlational structure. Hence, Figure \ref{fig:max hours} only shows a limited picture.

\subsection{Periods of positive residual load and heat deviation overlap sometimes}

\begin{figure}[!htb]
    \centering
        \begin{subfigure}{.45\textwidth}\centering\includegraphics[width=\textwidth]{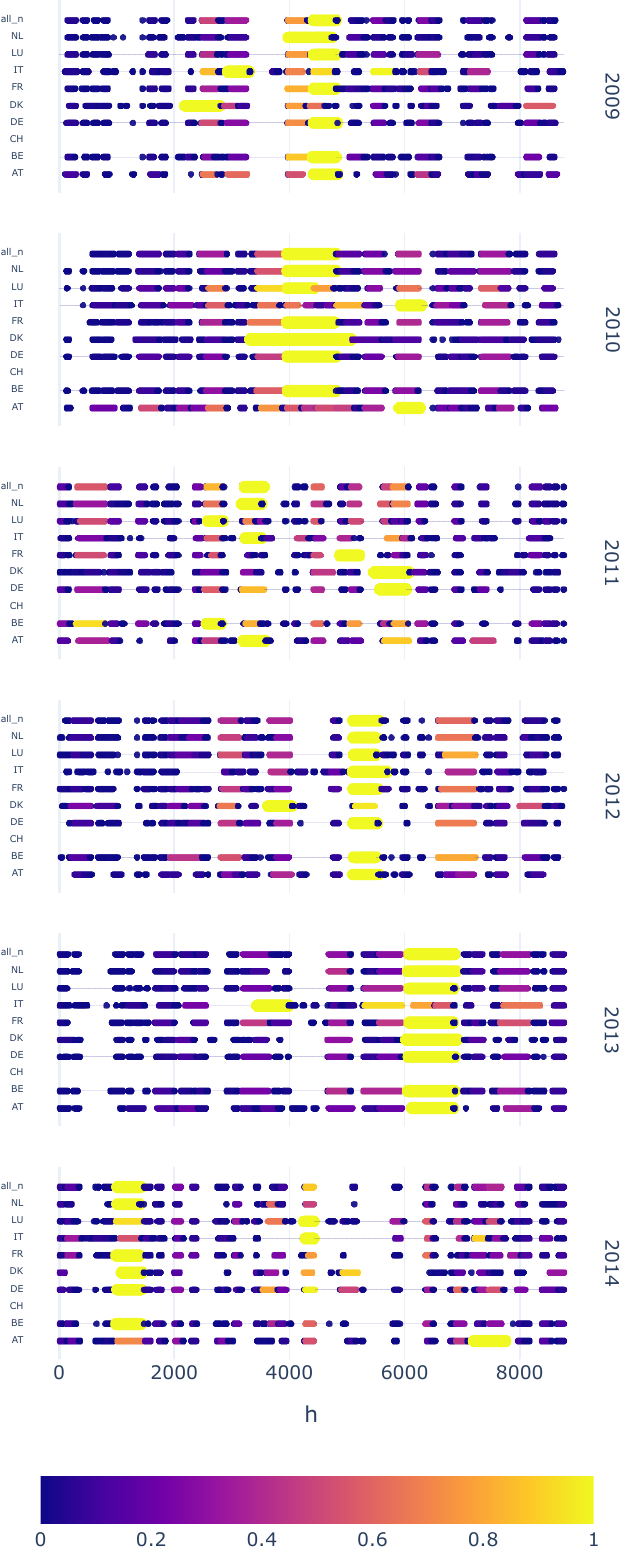}\subcaption[]{Maximum heat devition events}\label{fig:heat_demand_deviation}\end{subfigure}%
        \begin{subfigure}{.45\textwidth}\centering\includegraphics[width=\textwidth]{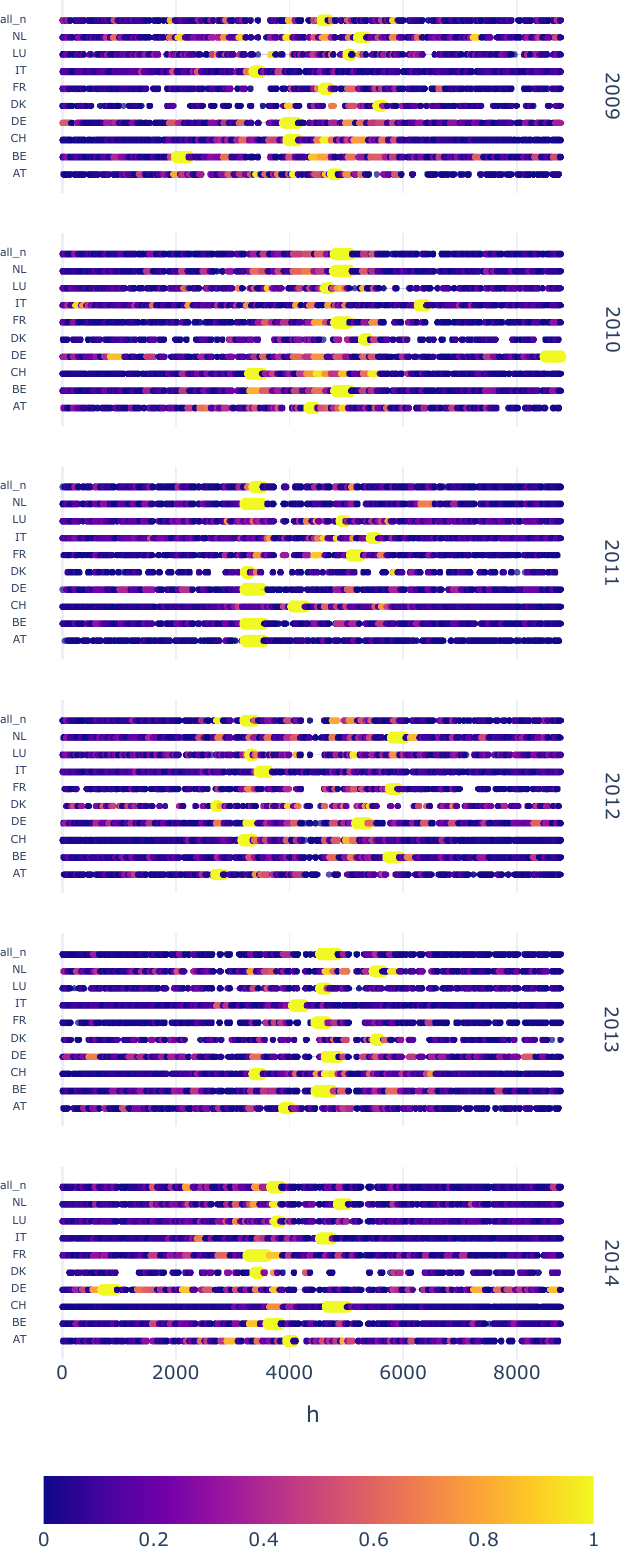}\subcaption[]{Maximum residual load event}\label{fig:residual_load_event}\end{subfigure}\par
        \begin{minipage}{0.9\textwidth}
           \medskip \footnotesize \textit{Note:} \textit{all\_n} refers to the sum of all countries.
        \end{minipage}\par
        \vspace{0.1cm}
    \caption{Maximum heat deviation and residual load events}
    \label{fig:max periods}
\end{figure}

However, the analysis of coincidence between the maximum heat pump load and residual load draws only a partial picture. The challenge for the energy system does not only arise from single hours of high (residual) load but also from more extended periods of low temperatures on the one side and longer periods of a positive residual load on the other side  (Figure \ref{fig:max periods}). If these two types of periods fall together, heat pumps add strain to the power sector. To analyze the occurrence of these two types of periods, the \textit{maximum heat deviation events}, defined as the cumulative sum of positive differences between the hourly heat demand and its average value are depicted (Figure \ref{fig:heat_demand_deviation}). When heat demand falls below its average value, a ``new'' event starts. Therefore, it is likely that the true length of the cold periods is underestimated. The difference from the average is a good indicator of ``cold spells'' creating possible difficulties for the power sector. For better visualization and comparability, the summed energy value of each event is related to the value of the maximum event in each country and year. The event with the largest positive heat demand deviation is labeled with a ``1'', and all the others, respectively, have values between zero and one depending on their relative size. Like in the analysis of hours of maximum heat demand, the maximum heat deviation events often align between countries, suggesting that they are driven by the same weather patterns. Please note, though, that the gravity of the events can be quite different between countries, as only the relative value to the maximum event of each country in each year is shown here. Years like 2010 and 2013 show how strongly heat deviation events can be correlated, suggesting that all countries were affected by the same weather events. Conversely, years like 2011 and 2014 depict shorter and less correlated events. Yet, maximum events in one country happen often in parallel to near-maximum events in other countries. With respect to positive residual load events, the picture looks a bit different (Figure \ref{fig:residual_load_event}). Periods of consecutive positive residual loads are shorter than heat deviation events. Yet, it is important to remember that the yellow events are only the biggest and are terminated when the residual load turns negative. Therefore, if not accounting for these short periods of negative residual load, one could define these positive residual load events even longer. 

Relevant to the power sector is whether there is an overlap between heat demand events and residual load events. Similarly to the previous analysis of maximum hours, the answer is mixed: while in some years, a clear overlap between the largest (and close-to-largest) residual load events and heat events (such as in 2010 and 2011) can be seen, in other years they do not fall together (such in 2014). This finding, combined with the insights drawn from Figure \ref{fig:max hours}, again shows the need to use several years in any energy system analysis to properly account for all possible phenomena and interactions. If the analysis of several years is not feasible, a careful study of weather years, which includes renewable energy availability and temperatures, is necessary to choose the appropriate year.
    
\subsection{Thermal energy storage reduces the need for firm generation capacity}

\begin{figure}[!htb]
    \centering
        \begin{subfigure}{.415\textwidth}\centering\includegraphics[width=\textwidth]{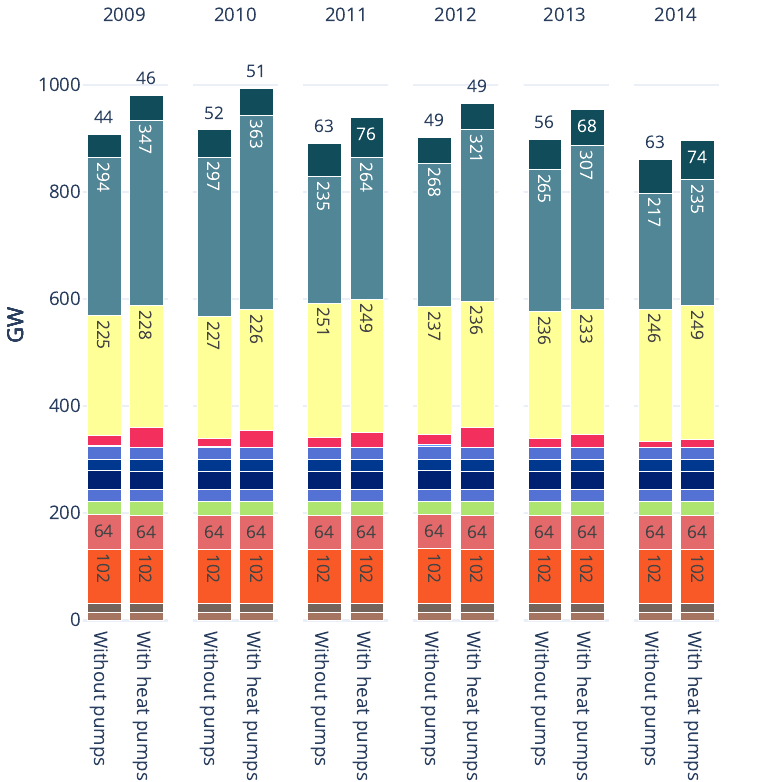}\subcaption[]{Two hour thermal storage}\label{fig:capacities 2h sto}\end{subfigure}%
        \begin{subfigure}{.58\textwidth}\centering\includegraphics[width=\textwidth]{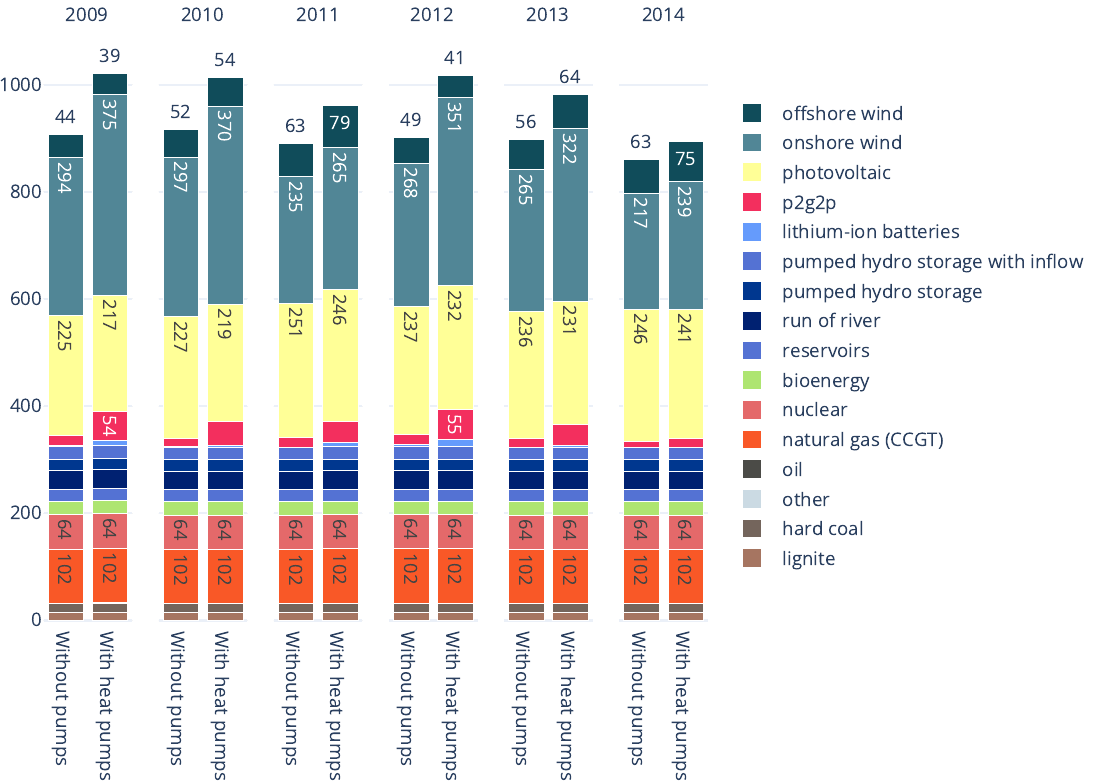}\subcaption[]{No thermal storage}\label{fig:capacities 0h sto}\end{subfigure}%
    \caption{Overview generation capacities}
    \label{fig:capacities overview}
\end{figure}

A substantial rollout of heat pumps requires additional electricity, hence additional power plant capacities. Figure \ref{fig:capacities overview} depicts total generation capacities in all countries, with and without heat pumps, for all weather years. Already in the scenarios without heat pumps, the electricity sector is rather wind-focused (217-297 GW), with capacity varying quite strongly between weather years. PV is also installed at a sizable dimension (225-246 GW). As assumed for the \textit{base} scenario, nuclear power, as well as coal (lignite and hard) are fixed, while gas-fired power has a lower bound. As the model chooses to keep the capacity of gas-fired power plants at that lower bound, it suggests that even lower capacities could be cost-optimal (see Section \ref{sec:robustness}). 

\begin{figure}[H]
    \centering
    \begin{subfigure}{.49\textwidth}\centering\includegraphics[width=\textwidth]{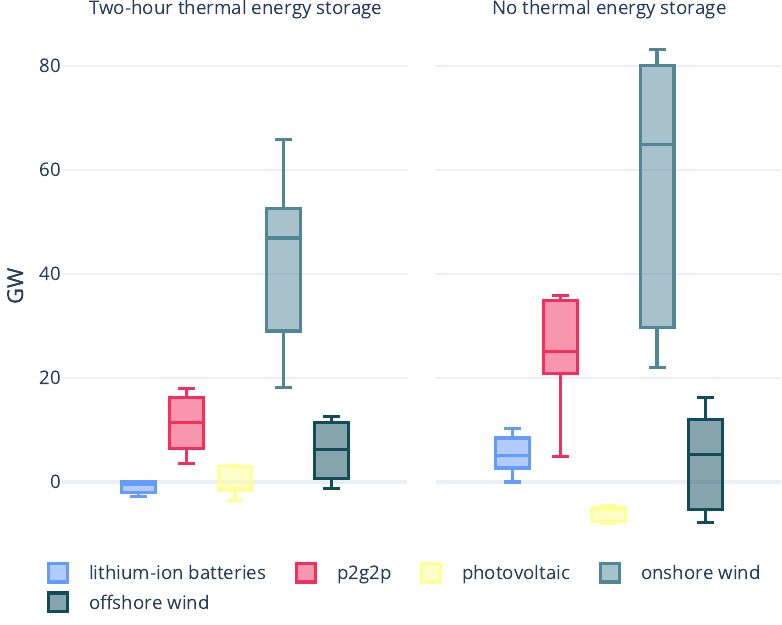}\subcaption[]{All capacities}\label{fig:main_capacities_aggregated_diff}\end{subfigure}%
    \begin{subfigure}{.49\textwidth}\centering\includegraphics[width=\textwidth]{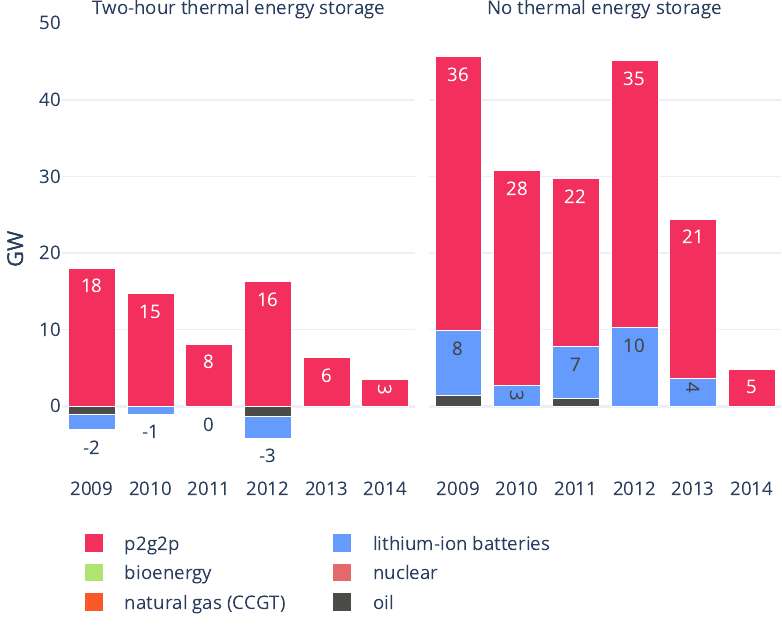}\subcaption[]{Firm capacities, per year}\label{fig:main_capacities_firm_diff}\end{subfigure}%
    \caption{Changes in aggregated generation capacities}
    \label{fig:capacities changes}
\end{figure}

The introduction of additional heat pumps leads primarily to more onshore wind power (20 to 80 GW) and \ac{p2g2p} storage (5 to almost 40 GW). Figure \ref{fig:capacities overview} shows the totals for every weather year, while Figure \ref{fig:capacities changes} depicts the changes as a box plot. P2g2p storage can be seen as a ``proxy'' for firm capacities. Interestingly, the model favors \ac{p2g2p} storage over expanding gas-fired power plants, likely because of high \ch{CO2} prices. Offshore wind power is also added in some years, yet at lower levels, due to the relatively high costs.

As shown in Figure \ref{fig:main_capacities_aggregated_diff}, equipping heat pumps with two-hour thermal energy storage leads to sizable differences in added capacities. While in the case of no thermal storage, the deployment of heat pumps leads to additional onshore wind capacities of between around 20 and 80 GW, these additional capacities are reduced to 20 to around 65 GW in the case of a two-hour thermal energy storage. Equally, the additions of \ac{p2g2p} and lithium-ion battery storage and offshore wind power are smaller or even negative. Depending on the weather year, the onshore wind power capacity is expanded between 10 and 30\%, while PV changes hardly.

Zooming in on only the firm capacities, the difference between the two scenarios is quite visible (Figure \ref{fig:main_capacities_firm_diff}). The two-hour thermal storage can avoid almost 20 GW of additional firm capacities, in my scenarios, mainly \ac{p2g2p} storage. If no thermal storage is allowed, sizable capacities of lithium-ion battery storage are added with the deployment of heat pumps, which add intraday flexibility. Allowing for the thermal energy storage, the additional lithium-ion batteries are not needed, as the thermal energy storage takes over that role, and even a reduction in lithium-ion batteries with the rollout of heat pumps can be seen.

\begin{figure}[htb]
    \centering
    \includegraphics[width=\textwidth]{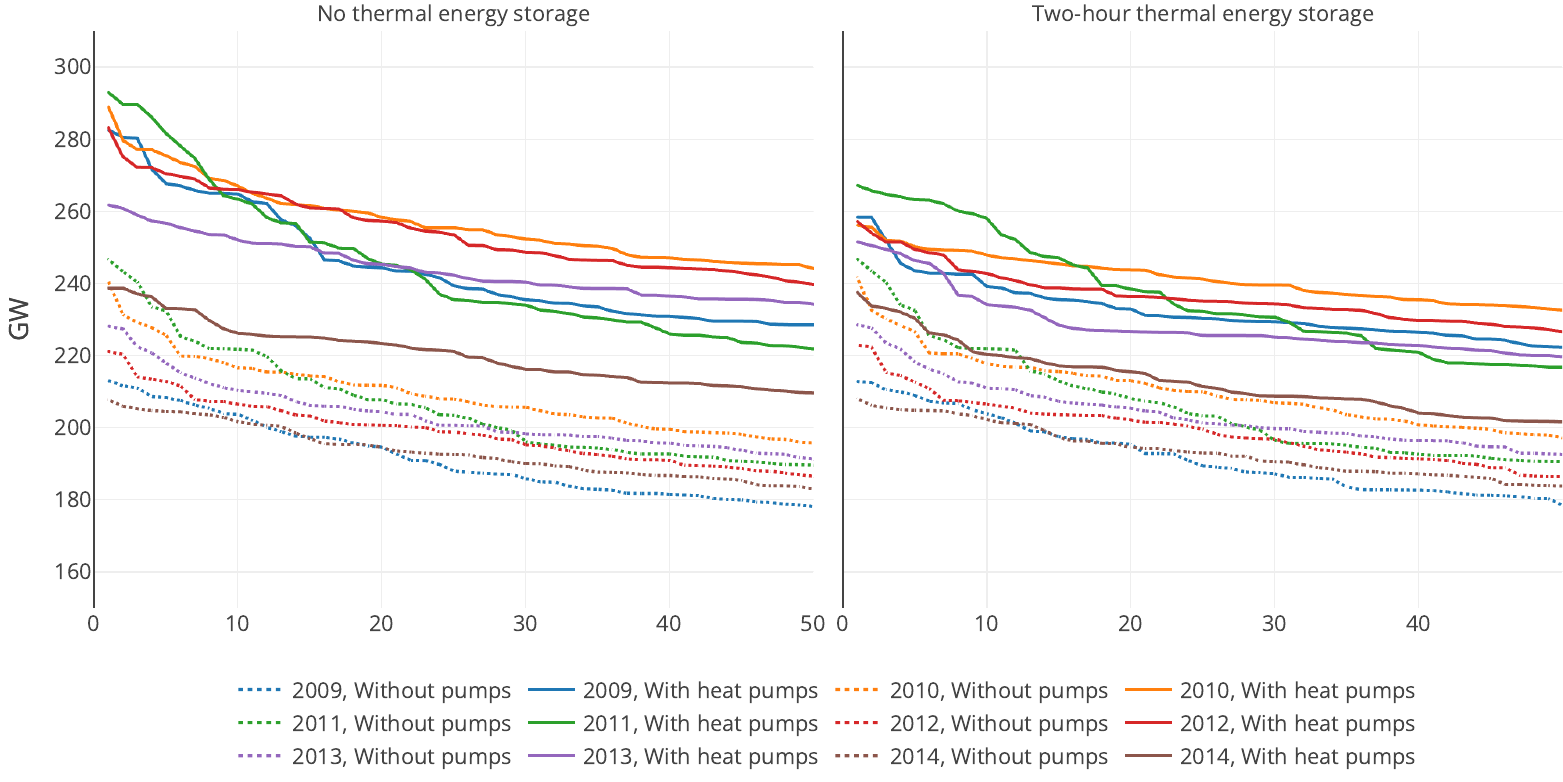}\par
    \begin{minipage}{\textwidth}
        \medskip \footnotesize \textit{Note:} The figure depicts the residual load duration curves of the ``system'', hence of all countries combined.
    \end{minipage}\par
    \vspace{0.1cm}
    \caption{Residual load duration curves}
    \label{fig:rldc ep}
\end{figure}

The effect of the thermal energy storage can also be nicely seen with the analysis of residual load duration curves (Figure \ref{fig:rldc ep}). The figure depicts the first 50 hours of the residual load duration curves of the entire ``system'', hence all countries. The dotted lines are the curves without the heat pump load, while the solid lines refer to the residual load duration curves with the heat pump load included. The left panel, the setting without thermal energy storage, reveals the direct impact of the heat pump load on the residual load. For instance, in the year 2009, there is a difference of almost 70 GW in residual load: heat pumps can add a lot of residual load! Not surprisingly, results differ quite substantially between years. The right panel displays the residual load duration curves in the scenario in which heat pumps are equipped with thermal energy storage of two hours. Compared to the left panel, the dotted lines, which are residual load duration curves without heat pump load, are almost the same. However, the solid lines are placed considerably lower, which shows that heat pumps in that scenario add considerably less residual load. For instance, in the year 2009, the difference between the two lines is considerably smaller, and the load added by heat pumps is now less than 50 GW. The effect visualized in this figure fits well with the results about additional firm generation capacities (Figure \ref{fig:main_capacities_firm_diff}), discussed above, and shows the importance of even — rather small — thermal storage for smoothing heating demand.

\subsection{Further results and robustness checks} \label{sec:robustness}

To better understand the quality of the results presented, a number of robustness checks are conducted in which some core assumptions of the model are varied. In the following, the generation capacity results and the impact on total system costs are presented and discussed. 

\begin{figure}[htb]
    \centering
    \begin{subfigure}{.5\textwidth}\centering\includegraphics[width=\textwidth]{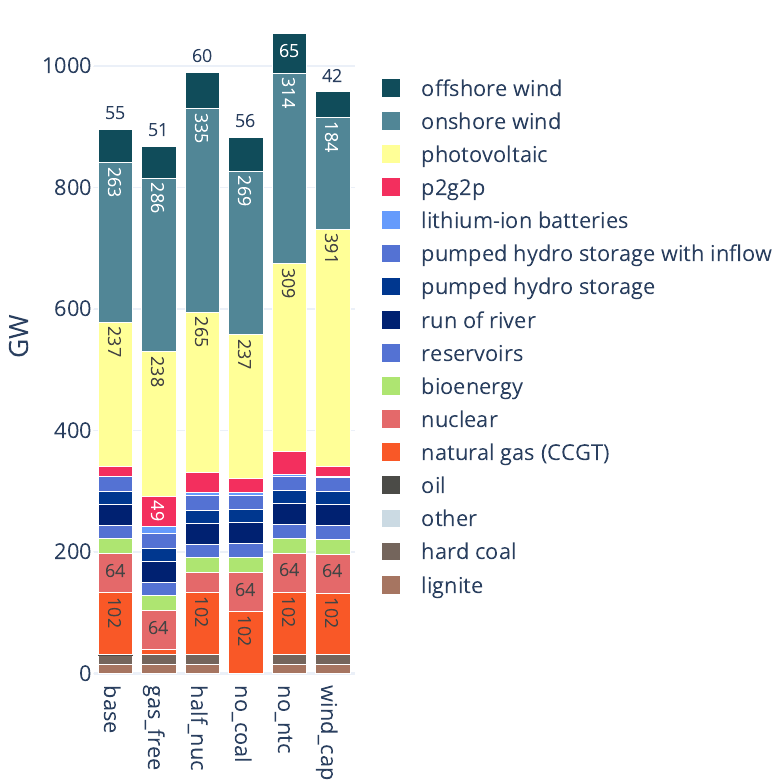}\subcaption[]{Average generation capacities without heat pumps}\label{subfig:main_capacities_rob_total}\end{subfigure}%
    \begin{subfigure}{.5\textwidth}\centering\includegraphics[width=\textwidth]{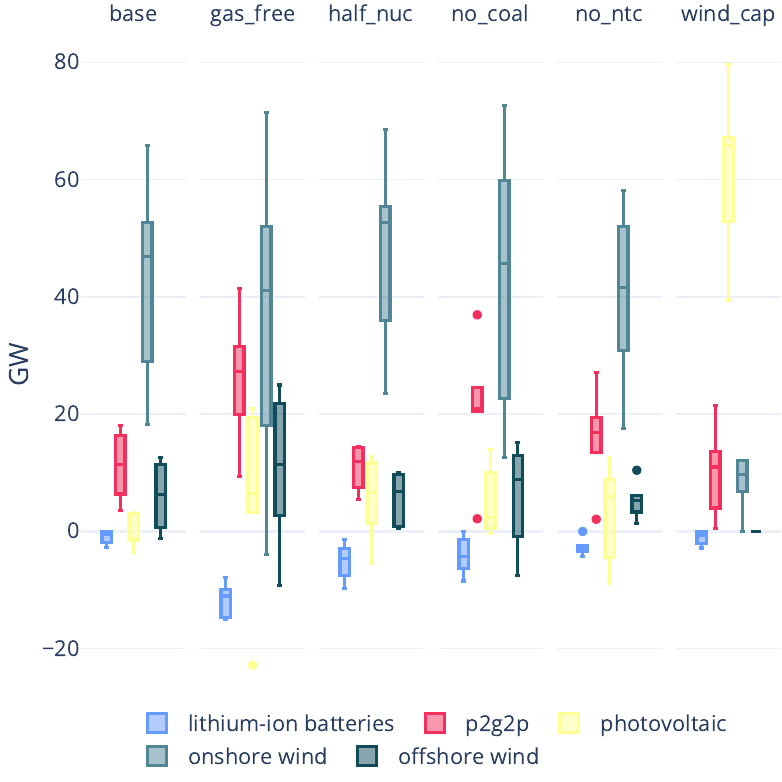}\subcaption[]{Capacity changes}\label{subfig:main_capacities_rob_diff}\end{subfigure}%
    \caption{Generation capacities in the robustness checks}
    \label{fig:capas robustness}
\end{figure}

Section \ref{sec:scenarios} describes the assumptions of the additional scenario runs. Please remember that all robustness scenarios (Table \ref{tab:robustness checks}) are conducted with heat pumps that have a two-hour thermal energy storage available. With respect to generation capacities, Figure \ref{fig:capas robustness} provides an overview of the capacities installed without heat pumps (\ref{subfig:main_capacities_rob_total}) and the changes due to heat pump deployment (\ref{subfig:main_capacities_rob_diff}). Several insights are paramount: if capacities of gas power plants are not fixed (scenario \textit{gas\_free}), the model chooses to remove them entirely and invests mainly in additional \ac{p2g2p} storage and onshore wind plants (compared to scenario \textit{base}). That effect can be explained by the high \ch{CO2}, which renders, in turn, the operation of gas-power power plants costly. Regarding p2g2p storage, the scenarios \textit{half\_nuc}, \textit{no\_coal}, and \textit{no\_ntc} foresee higher investments to replace either missing firm capacity or due to the removed flexibility without electricity exchange. The scenarios \textit{half\_nuc} and \textit{no\_ntc} foresee considerably higher capacities of onshore wind power to replace missing energy generation and trade. The scenario \textit{no\_ntc} also foresees considerably higher PV capacities, not surprising as every country has to work in autonomy and therefore requires an overall more balanced power plant portfolio. Finally, the scenario \textit{wind\_cap} leads to very high capacities of PV compared to \textit{base}, to replace the electricity formerly generated by onshore wind power.

The deployment of heat pumps has different effects on installed generation and storage discharge capacities (Figure \ref{subfig:main_capacities_rob_diff}). Compared to the \textit{base} scenario, additional on- and offshore wind power capacities are relatively similar, which is also not too surprising given the restrictions of the model and the high \ch{CO2} price: (onshore) wind power is the most cost-effective technology to provide the additional electricity needed. In the scenario \textit{wind\_cap}, the heat pump rollout leads to mainly additional PV capacity, as on- and offshore wind capacities are already close to or at their respective upper bounds. Regarding the firm capacities, comparable dynamics can be seen in most scenarios, and several insights can be drawn: the additional heat pumps require additional firm capacities, which are provided by the p2g2p storage. Depending on the year, the assumed heat pump rollout can require between a few and almost 20 GW of additional p2g2p storage (in the \textit{base} scenario). With higher capacities of p2g2p storage, lithium-ion batteries are pushed out of the system. In a system without interconnection (scenario \textit{no\_ntc}), the additional p2g2p capacities are similar to the ones in \textit{base}, suggesting that interconnection provides only little additional flexibility to cope with the heat pump load. This aligns with the insights from the previous section \ref{sec:results heat}, which demonstrated that heat demand peaks and strong heating periods overlapped between many countries. The two scenarios \textit{gas\_free} and \textit{no\_coal} trigger unsurprisingly additional investments into p2g2p storage to replace the missing firm capacities. The scenario \textit{half\_nuc} is similar to \textit{base}, suggesting that the nuclear power plants (mostly in France) are indeed not crucial in delivering firm capacities to cover peak loads from heat pumps.

\begin{figure}[H]
    \includegraphics[width=\linewidth]{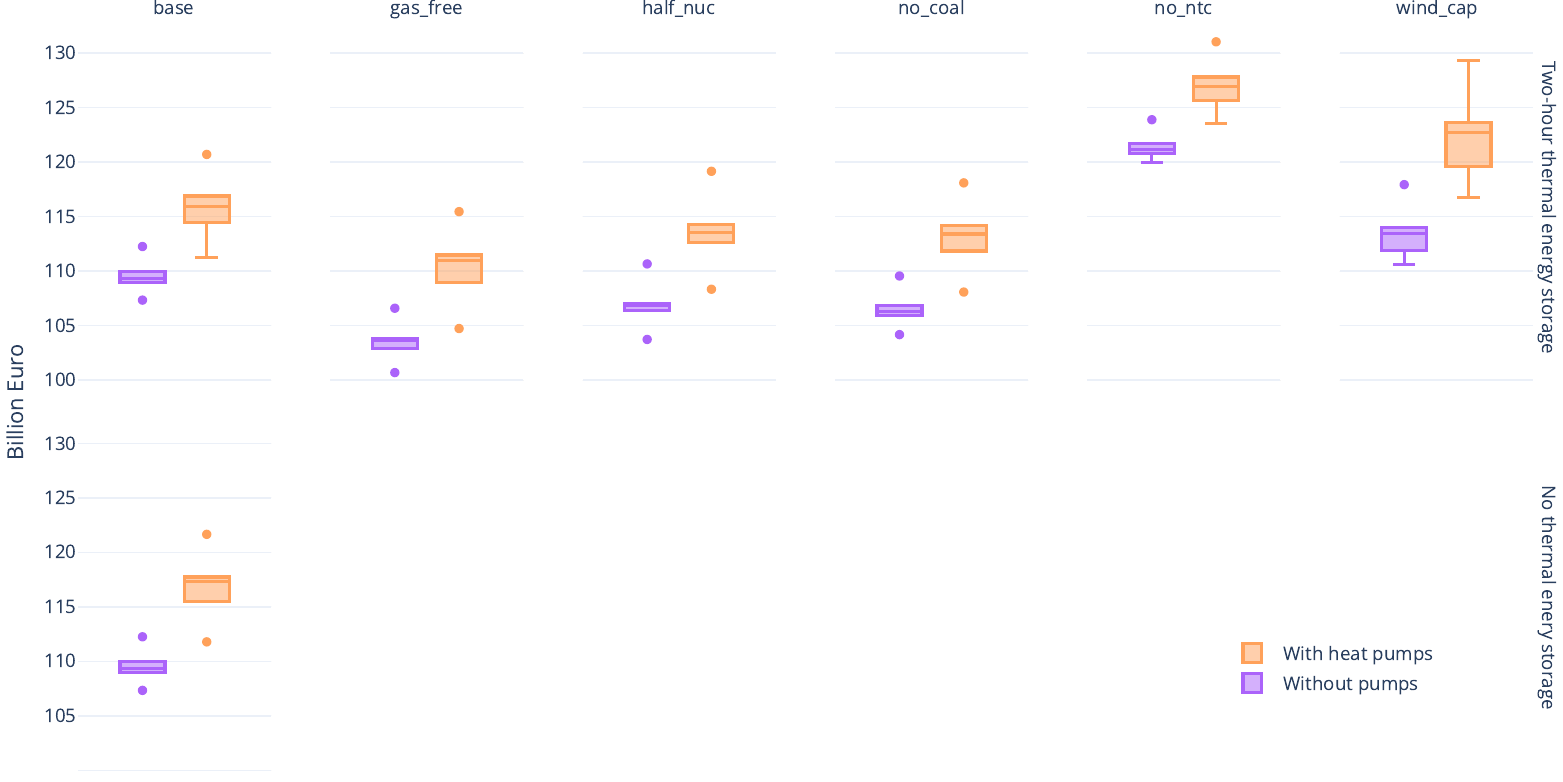}
    \caption{Total system costs}
    \label{fig:total system cost}
\end{figure}

With respect to the different scenarios, total system costs are similar in the scenarios \textit{base}, \textit{half\_nuc}, and \textit{no\_coal}, with even lower values in the latter two (Figure \ref{fig:total system cost}). The most expensive scenarios are \textit{no\_ntc} and \textit{wind\_cap}. Especially in the latter, the additional heat pumps increase system costs considerably. In the former, total system costs are on a higher overall level, but the introduction of heat pumps leads to a similar cost increase as in the \textit{base} scenario, suggesting that an interconnected system does not provide much flexibility to cover the additional heat pump load. For the \textit{wind\_cap} scenario, costs without heat pumps are not much higher compared to the scenario \textit{base}, yet increase substantially with the heat pump rollout, showing the compatibility of (onshore) wind power and heat pumps \citep[cf][]{ruhnau_heating_2020}. The impact of thermal heat storage on total system costs is small. Finally, Figure \ref{fig:total system cost} shows again the variability of results with respect to weather years. Consistently, the year 2010 constitutes an upper outlier, while 2014 is a lower outlier. Further figures on the total electricity generation (Figure \ref{fig:generation}) and on all residual load duration curves (Figure \ref{fig:all rldc}) can be found in the appendix. 

The introduction of heat pumps leads to an increase in total system costs of around 5 billion EUR, as additional generation capacities need to be installed and additional electricity generated. These costs just reflect the costs of the electricity sector and do not account for the costs of heat pumps, for instance. As shown in Figure \ref{fig:heat_demand_total}, total heat demand in my system ranges around 1,300 TWh. If heat pumps would cover 25\% of that demand, they would supply around 325 TWh of heat, translating to a price of 15.5 EUR/MWh. That price level is well below current wholesale prices of natural gas (around 40 EUR/MWh at the time of writing), assuming that the heat supplied with heat pumps had been previously generated by only burning natural gas with an efficiency of 100\%. This calculation does not even account for the \ch{CO2} price to be added to the gas bill. Therefore, it is evident, at least in terms of variable costs, that the additional expenses for the electricity sector costs are very favorable compared to the expenses for natural gas needed to generate the same amount of heat.

\subsection{Limitations}

As in any model-based analysis, my results depend strongly on data and scenario assumptions. Regarding the modeling of heat, I make several simplifying assumptions such that I do not model the detailed physical properties of heat pumps, only considering one heat pump technology, and also abstracting from potential flexibility by existing heat pumps as I only model additional units. Also, I exogenously set the share of heat covered by heat pumps and assume the size of the thermal energy storage instead of determining these variables endogenously. I abstract from any market incentives for consumers to operate heat pumps in a system-optimal way but assume they act in a system-friendly manner. My analysis could also be improved by adding more countries and weather years. Finally, I abstract from any detailed transmission and distribution grid modeling.

\section{Conclusion}

Heat pumps are a cornerstone in decarbonizing the heat supply of European buildings. Yet, their deployment does not come without challenges for the power sector. The present analysis evaluates a simultaneous heat pump rollout in several European countries, which allows several conclusions to be drawn. First, an ambitious rollout requires the installation of additional electricity generation capacities. Covering 25\% of total heat demand in buildings by air-sourced heat pumps would require around 50 GW of additional onshore wind power capacity, alongside additional storage and firm capacities of far lower magnitude. In the case of expansion limits of onshore wind, the additional electricity could also be supplied by solar PV in combination with storage. Second, the flexibility of heat pumps is critical. Even a small thermal energy storage in combination with a system-friendly operation of heat pumps leads to a sizable reduction in peak loads and, therefore, firm capacity needs. Third, the interconnection between countries does not substantially help to reduce generation (and firm) capacities, as cold spell events are correlated. Hence, fourth, it is paramount to understand properly the nexus of heat demand and renewable energy supply in order to adequately assess the challenges of a further electrification of heat. Fifth, the additional costs of the electricity sector are very favorable compared to expenses for natural gas to generate a similar amount of heat. Finally, the present analysis shows again the variability of results with respect to different weather years. Thus, the statements of policy-informing studies should be interpreted with the insight in mind that results might strongly vary depending on weather data. The choice of the ``right'' weather year, the usage of multiple weather years, or even better, the modeling of multi-year periods periods are possible ways forward.

\clearpage

\newpage

\printbibliography

\newpage

\appendix
\section*{Supplemental information}

\subsection*{Capacity bounds, costs, and technical parameters}

\begin{table}[H]

\centering
\caption{Cost and technology parameters}
\label{tab: costs}

\begin{subtable}[t]{\textwidth}
\caption{Electricity storage and reservoirs}
\tiny
\setlength{\tabcolsep}{2pt}
\begin{tabulary}{\textwidth}{L|CCC|CCC|CC|CC}
\toprule
                                                &
  \textbf{Interest rates}                       &
  \textbf{Lifetime}                             &
  \textbf{Availability}                         &
  \multicolumn{3}{c|}{\textbf{Overnight costs}} &
  \multicolumn{2}{c|}{\textbf{Efficiency}}      &
  \multicolumn{2}{c}{\textbf{Marginal costs}} 
  \\
  
  &
  &
  & 
  &
  \textbf{energy} &
  \textbf{charging power} &
  \textbf{discharging power} &
  \textbf{charging} &
  \textbf{discharging} &
  \textbf{charging} &
  \textbf{discharging}
  \\
  \textbf{Technology} &
  [years]             &
  [years]             &
  [years]             &
  [1000 EUR]          &
  [1000 EUR]          &
  [1000 EUR]          &
                      &
                      &
       [EUR]          &
       [EUR]          \\ \midrule
Lithium-ion batteries      & \multirow{4}{*}{0.04} & 20 & 0.98 & 300 & 50  & 10  & 0.97 & 0.97 & 0.3  & 0.3 \\
Power-to-gas-to-power      &                       & 23 & 0.95 & 0.2 & 305 & 850 & 0.73 & 0.6  & 1.2  & 1.2 \\
Pumped hydro (open/closed) &                       & 80 & 0.98 & 10  & 550 & 550 & 0.97 & 0.91 & 0.56 & 0.56 \\
Hydro reservoirs           &                       & 50 & 0.98 & 10  & 200 & -   & 1.00 & 0.95 & 0    & 0.1 \\
\bottomrule
\end{tabulary}%
\end{subtable}

\vspace{1cm}

\begin{subtable}[t]{\textwidth}
\caption{Electricity generation}

\tiny

\centering

\begin{tabularx}{\textwidth}{l|cccccccc}
\toprule
  \textbf{Technology} &
  \textbf{Interest rates} &
  \textbf{Lifetime} &
  \textbf{Availability} &
  \textbf{Overnight costs} &
  \textbf{Fixed costs} &
  \textbf{Efficiency} &
  \textbf{Carbon content} & 
  \textbf{Fuel costs}  
  \\
                              &
                              &
  [years]                     &
                              &
  [1000 EUR]                  &
  [1000 EUR]                  &
                              & 
  [t/MWh]   &
  [EUR/MWh]  \\ \midrule
Closed-cycle gas turbine & \multirow{11}{*}{0.04} & 25 & 0.96 & 830   & 28 & 0.61 & 0.20 & 26.0 \\
Bioenergy                &                        & 25 & 1.00 & 900   & 9  & 0.45 & 0.00 & 10.0 \\
Hard coal                &                        & 35 & 0.96 & 1,300 & 30 & 0.43 & 0.34 & 10.1 \\
Lignite                  &                        & 35 & 0.95 & 1,500 & 30 & 0.38 & 0.40 & 4.0 \\
Nuclear                  &                        & 40 & 0.91 & 6,000 & 30 & 0.34 & 0.00 & 1.7 \\
Oil                      &                        & 25 & 0.90 & 400   & 7  & 0.35 & 0.27 & 41.7 \\
Other                    &                        & 30 & 0.90 & 1,500 & 30 & 0.35 & 0.35 & 18.1 \\
Solar photovoltaic       &                        & 40 & 1.00 & 597   & 10 & 1.00 & 0.00 & 0.0 \\
Wind onshore             &                        & 50 & 1.00 & 3,000 & 30 & 0.90 & 0.00 & 0.0 \\
Wind offshore            &                        & 30 & 1.00 & 1,795 & 39 & 1.00 & 0.00 & 0.0 \\
Run-of-river             &                        & 30 & 1.00 & 1,036 & 13 & 1.00 & 0.00 & 0.0 \\
\bottomrule

\end{tabularx}%

\end{subtable}

\end{table}

\begin{table}[H]
\centering
\caption{Assumptions on capacity bounds [in GW]}
\label{tab:capacity_bounds}%
  
\resizebox{\textwidth}{!}{
\begin{tabular}{l|cccccccccccccccccc}
    \toprule
    \textbf{Technology} & \multicolumn{2}{c}{\textbf{Austria}} & \multicolumn{2}{c}{\textbf{Belgium}} & \multicolumn{2}{c}{\textbf{Denmark}} & \multicolumn{2}{c}{\textbf{France}} & \multicolumn{2}{c}{\textbf{Germany}} & \multicolumn{2}{c}{\textbf{Italy}} & \multicolumn{2}{c}{\textbf{Luxembourg}} & \multicolumn{2}{c}{\textbf{Netherlands}} & \multicolumn{2}{c}{\textbf{Switzerland}}  \\
                                  & low     & up    & low     & up  & low     & up  & low     & up  & low     & up   & low     & up & low     & up     & low     & up       & low     & up \\
    \midrule
    Natural gas (CCGT)            & 4.0     & inf   & 8.1     & inf & 4.0     & inf & 7.2    & inf  & 25.4    & inf  & 40.5  & inf  & 0          & inf & 12.4        & inf  & 0           & inf   \\
    Oil                           & 0       & 0.16  & 0       & 0.2 & 0       & 2.5 & 0      & 1.3  & 0       & 1.0  & 0     & 0    & 0          & 0   & 0           & 0    & 0           & 0     \\
    Other                         & 0       & 0.96  & 0       & 1.4 & 0       & 1.3 & 0      & 5.7  & 0       & 8.8  & 0     & 6.4  & 0          & 0.1 & 0           & 4.2  & 0           & 0.6   \\
    Hard coal                     & 0       & 0     & 0       & 0   & 1.2     & 1.2 & 0      & 0    & 12.3    & 12.3 & 0     & 0    & 0          & 0   & 2.7         & 2.7  & 0           & 0     \\
    Lignite                       & 0       & 0     & 0       & 0   & 0       & 0   & 0      & 0    & 14.6    & 14.5 & 0     & 0    & 0          & 0   & 0           & 0    & 0           & 0     \\
    Nuclear                       & 0       & 0     & 0       & 0   & 0       & 0   & 61.8   & 61.8 & 0       & 0    & 0     & 0    & 0          & 0   & 0.5         & 0.5  & 2.2         & 2.2   \\
    Bioenergy                     & 0.6     & 0.6   & 0.9     & 0.9 & 6.8     & 6.8 & 2.3    & 2.3  & 7.2     & 7.2  & 4.5   & 4.5  & 0.08       & 0.08 & 1.9        & 1.9  & 0.4         & 0.4   \\
    Run-of-river hydro            & 6.1     & 6.1   & 0.1     & 0.1 & 0       & 0   & 13.6   & 13.6 & 4.7     & 4.7  & 6.2   & 6.2  & 0.04       & 38  & 0.04        & 0.04 & 4.2         & 4.2   \\
    Solar PV                      & 5.0     & inf   & 7.5     & inf & 15.4    & inf & 18.2   & inf  & 74.5    & inf  & 28.6  & inf  & 0.3        & inf & 18.7        & inf  & 5.5         & inf   \\
    Onshore wind                  & 5.5     & inf   & 3.6     & inf & 16.4    & inf & 24.1   & inf  & 64.0    & inf  & 15.7  & inf  & 0.3        & inf & 6.0         & inf  & 0.2         & inf   \\
    Offshore wind                 & 0       & inf   & 2.3     & inf & 10.0    & inf & 2.5    & inf  & 11.1    & inf  & 0.3   & inf  & 0          & inf & 5.9         & inf  & 0           & inf   \\
    Lithium-ion batteries         &         &       &         &     &         &     &        &      &         &      &       &      &            &     &             &      &             &       \\
    ... power in/out              & 0       & inf   & 0       & inf & 0       & inf & 0      & inf  & 0       & inf  & 0     & inf  & 0          & inf & 0           & inf  & 0           & inf   \\
    ... energy [GWh]              & 0       & inf   & 0       & inf & 0       & inf & 0      & inf  & 0       & inf  & 0     & inf  & 0          & inf & 0           & inf  & 0           & inf   \\
    Power-to-gas-to-power         &         &       &         &     &         &     &        &      &         &      &       &      &            &     &             &      &             &       \\
    ... power in/out              & 0       & inf   & 0       & inf & 0       & inf & 0      & inf  & 0       & inf  & 0     & inf  & 0          & inf & 0           & inf  & 0           & inf   \\
    ... energy [GWh]              & 0       & inf   & 0       & inf & 0       & inf & 0      & inf  & 0       & inf  & 0     & inf  & 0          & inf & 0           & inf  & 0           & inf   \\
    Pumped hydro storage (closed) &         &       &         &     &         &     &        &      &         &      &       &      &            &     &             &      &             &       \\
    ... power in                  & 0.3     & 0.3   & 1.2     & 1.2 & 0       & 0   & 2.0    & 2.0  & 7.4     & 7.4  & 7.4   & 7.4  & 1.0        & 1.0 & 0           & 0    & 1.9         & 1.9   \\
    ... power out                 & 0.3     & 0.3   & 1.2     & 1.2 & 0       & 0   & 2.0    & 2.0  & 7.4     & 7.4  & 7.3   & 7.3  & 1.3        & 1.3 & 0           & 0    & 1.9         & 1.9   \\
    ... energy [GWh]              & 1.8     & 1.8   & 5.3     & 5.3 & 0       & 0   & 10     & 10   & 242     & 242  & 70.4  & 70.4 & 5.0        & 5.0 & 0           & 0    & 70          & 70    \\
    Pumped hydro storage (open)   &         &       &         &     &         &     &        &      &         &      &       &      &            &     &             &      &             &       \\
    ... power in                  & 5.2     & 5.2   & 0       & 0   & 0       & 0   & 1.9    & 1.9  & 1.4     & 1.4  & 2.1   & 2.1  & 0          & 0   & 0           & 0    & 2.1         & 2.1   \\
    ... power out                 & 6.0     & 6.0   & 0       & 0   & 0       & 0   & 1.9    & 1.9  & 1.6     & 1.6  & 3.3   & 3.3  & 0          & 0   & 0           & 0    & 10.7        & 10.7  \\
    ... energy [GWh]              & 1,732   & 1,732 & 0       & 0   & 0       & 0   & 90     & 90   & 417     & 417  & 309   & 309  & 0          & 0   & 0           & 0    & 8,800       & 8,800 \\
    Reservoirs                    &         &       &         &     &         &     &        &      &         &      &       &      &            &     &             &      &             &       \\
    ... power out                 & 2.5     & 2.5   & 0       & 0   & 0       & 0   & 8.9    & 8.9  & 1.3     & 1.3  & 9.6   & 9.6  & 0          & 0   & 0           & 0    & 0           & 0     \\
    ... energy [TWh]              & 0.8     & 0.8   & 0       & 0   & 0       & 0   & 10     & 10   & 0.2     & 0.2  & 5.6   & 5.6  & 0          & 0   & 0           & 0    & 0           & 0     \\
    \bottomrule
\end{tabular}%
}

\end{table}

\subsection*{Results}

\begin{figure}[H]
    \centering
    \includegraphics[width=1\linewidth]{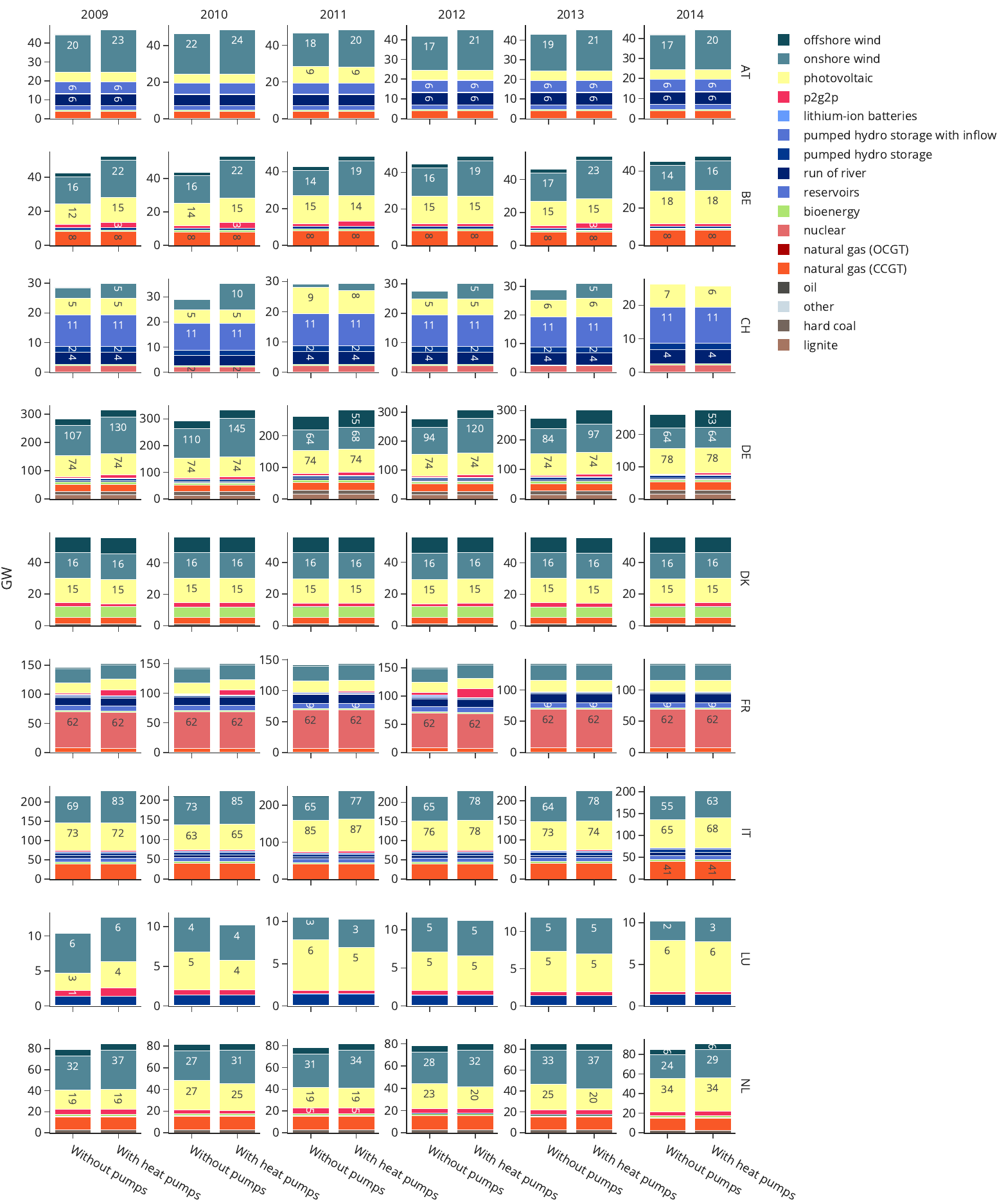}
    \caption{Generation capacities of all countries}
    \label{fig:capas all countries}
\end{figure}

Figure \ref{fig:capas all countries} depicts the generation capacities of all countries in all assessed weather years for the \textit{base} scenarios with thermal heat storage of 2 hours.

\begin{figure}[H]
    \includegraphics[width=\linewidth]{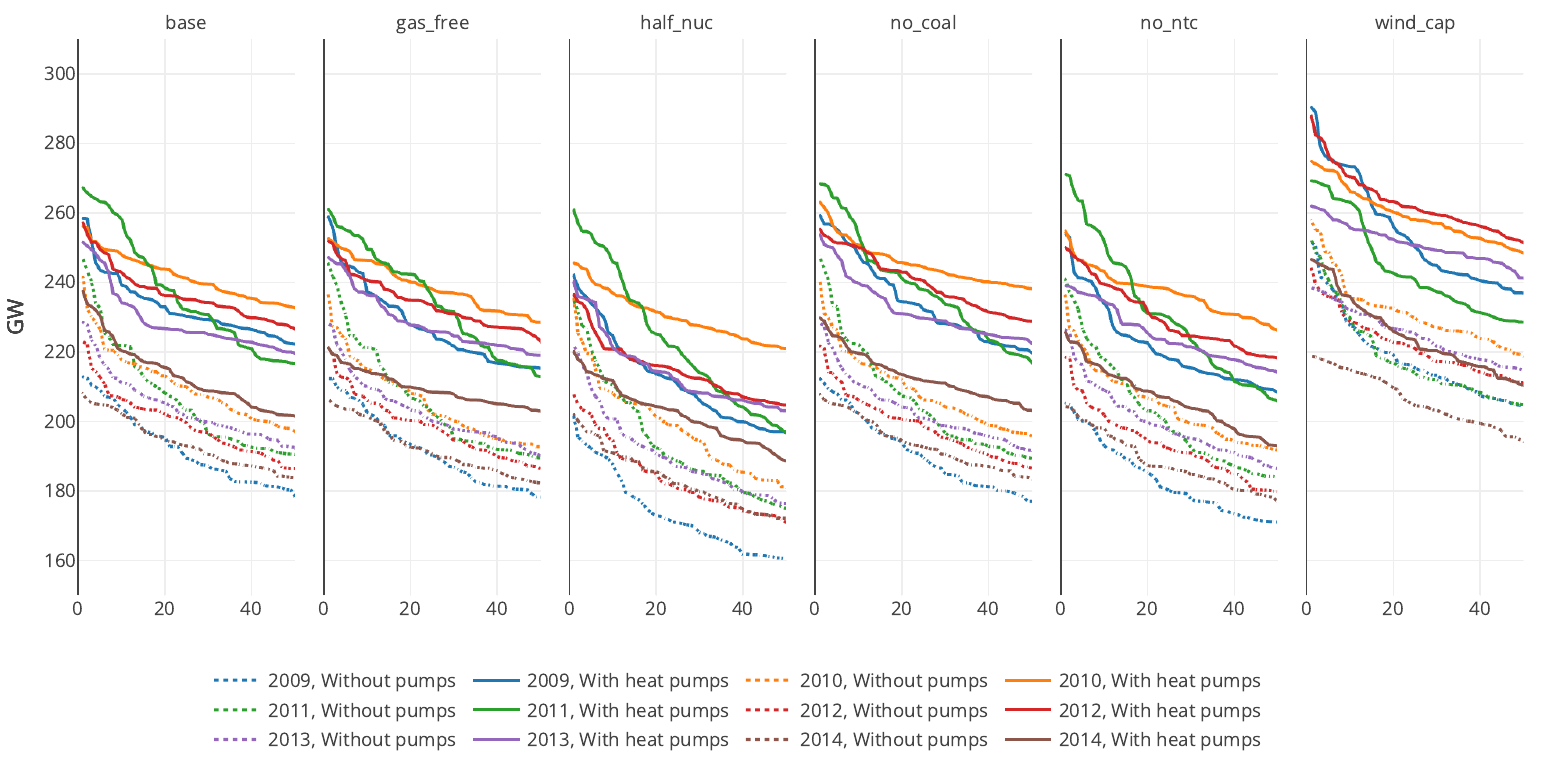}
    \caption{All residual load curves}
    \label{fig:all rldc}
\end{figure}

Figure \ref{fig:all rldc} depicts the residual load duration curves in all scenarios and all years, with and without a heat pump rollout. The jump caused by heat pumps is clearly visible, showing itself in the differences in firm generation capacities (see Figure \ref{fig:capas robustness}). While the patterns are relatively similar in all scenarios, \textit{wind\_cap} shows a clearly higher level as the missing wind power pushes up the residual load duration curve.

\begin{figure}[H]
    \includegraphics[width=\linewidth]{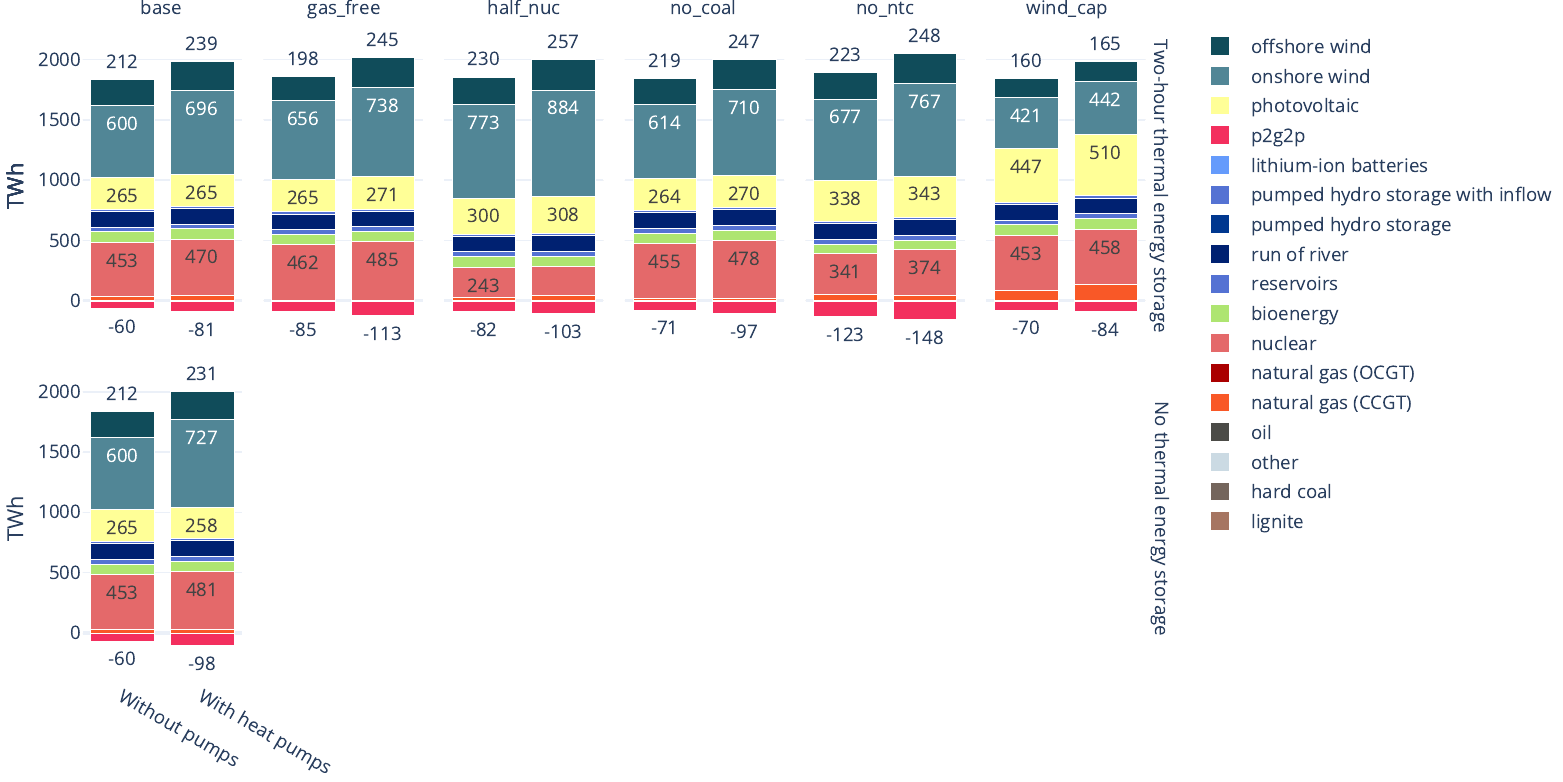}
    \caption{Electricity generation (average all years)}
    \label{fig:generation}
\end{figure}

Figure \ref{fig:generation} shows total electricity generation by technology as an average over all years. Storage technologies have negative values as discharging minus charging is shown, considering inefficiencies.

\end{document}

%% file: tables/table_cap.tex
\begin{tabular}{lrrr}
\toprule
Country &  Heat output (GW$_{th}$) &  Heat storage (GWh$_{th}$) &  Electricity input (GW$_{el}$) \\
\midrule
     AT &                      5.5 &                       11.0 &                            3.5 \\
     BE &                      8.8 &                       17.7 &                            5.1 \\
     CH &                      0.0 &                        0.0 &                            0.0 \\
     DE &                     63.8 &                      127.5 &                           39.7 \\
     DK &                      3.9 &                        7.8 &                            1.9 \\
     FR &                     41.1 &                       82.2 &                           20.9 \\
     IT &                     29.2 &                       58.4 &                           13.9 \\
     LU &                      0.7 &                        1.3 &                            0.4 \\
     NL &                     12.5 &                       25.0 &                            6.8 \\
 \hline   All &                    165.4 &                      330.9 &                           92.0 \\
\bottomrule
\end{tabular}